\documentclass[12pt]{article}
\usepackage{color}
\usepackage{amsmath}
\usepackage{hyperref}
\usepackage[T1]{fontenc}
\usepackage{epsfig}
\usepackage{hhline}
\usepackage{amssymb}
\usepackage{times}
\usepackage{cite}

\newlength{\dinwidth}
\newlength{\dinmargin}
\begin{document}  

\def\Journal#1#2#3#4{{#1}~{\bf #2} (#3) #4}
\def\EJC{{\em Eur. Phys. J.} {\bf C}}

\begin{titlepage}

\noindent
\begin{flushleft}
{\tt DESY 12-062    \hfill    ISSN 0418-9833} \\
{\tt 
August 2013}                  \\
\end{flushleft}

\noindent

\vspace{2cm}
\begin{center}
\begin{Large}
Erratum to \\
{\bf Determination of the Integrated Luminosity at HERA using Elastic
  QED Compton Events} \\
Eur. Phys. J. C72 (2012) 2163 

\vspace{2cm}

H1 Collaboration

\end{Large}
\end{center}

\vspace{2cm}


\vspace{1.5cm}

\begin{center}
Submitted to \EJC{}
\end{center}

\end{titlepage}


The calculation of the QED Compton (QEDC) cross sections in a recent
H1 publication \cite{Aaron:2012kn} is based on the COMPTON22 event
generator \cite{compton22_1}.
The cross sections for elastic, quasi-elastic and inelastic QEDC 
are all proportional to $(\hbar c)^2\alpha^3/s$, where
$\alpha$ is the fine-structure constant, $s$ is the centre-of-mass
energy squared measured in $\text{GeV}^2$ and the factor $(\hbar c)^2$
is a conversion factor, in units of $[\text{pb}\,\text{GeV}^2]$.
It was found that incorrect numerical values have been programmed in
the FORTRAN code used to calculate the cross sections, namely 
$(\hbar c)^2=(6.20087)^2\times10^7\,\text{pb}\,\text{GeV}^2$
and $\alpha=1/137$.
These numbers are corrected to $(\hbar c)^2=0.389379\times
10^9\,\text{pb}\,\text{GeV}^2$ and $\alpha=0.00729735$, thus
enhancing the predicted QEDC cross sections by $1.19\%$.

Furthermore, the running of the electromagnetic coupling as a function
of the virtuality $t$ of the exchanged photon is neglected in
\cite{compton22_1}.
For the reanalysis of \cite{Aaron:2012kn}, a running fine-structure constant
$\alpha(t)$ is implemented.
The cross sections predicted by COMPTON22 are scaled by a factor
$(\alpha(t)/\alpha)^2$ prior to integrating over $t$.
The running coupling $\alpha(t)$ is evaluated using the {\tt alphaQED}
code \cite{Jegerlehner:2011mw}.
This change increases the predicted cross sections in the analysis
phase space \cite{Aaron:2012kn} by $(0.83\pm0.04)\%$, where the
uncertainty is related to the number of generated events.

Taking both effects together, the predicted elastic QEDC cross
section in the analysis phase space is increased by $2.0\%$.
The corrected cross sections in the generated phase space is
$\sigma_{\text{gen}}=55.9\,\text{pb}$, where the estimated uncertainty
on the QEDC theory of $1.1\%$ is unchanged.
Similarly, the visible cross section is increased to
$\sigma_{\text{vis}}=37.1\,\text{pb}$.
The background fractions also change slightly, as shown in Table
\ref{table:background}.
Background from quasi-elastic and inelastic
QEDC processes increases, because the $t$ dependence is different from
that of the elastic QEDC process.
The relative fractions of the other background sources
are reduced, because their predicted absolute cross
sections do not change.

In addition to the above changes in the cross section prediction, a
small inefficiency in the data handling has been identified.
After correcting this technical problem, $21$ additional
data events are recovered for the luminosity measurement, now
derived from a total of $14298$ candidate events.

The overall HERA luminosity in the data taking period from 2003 to
2007, measured from counting QED Compton events, is found to be $345.3\pm
7.9\,\text{pb}^{-1}$.
As compared to \cite{Aaron:2012kn} it is lower by $1.8\%$.
The cross section measurements performed in three H1 papers 
\cite{Aaron:2012cj,Aaron:2012qi,Alexa:2013xxa}, based on the data
collected in the years 2003-2007 at a proton energy of
$920\,\text{GeV}$,  are normalised using the integrated
luminosity measurement of \cite{Aaron:2012kn}.
For this reason they are affected by the change in the measured
luminosity discussed above, such that their cross sections are to be
scaled up by $1.8\%$. It is worth to note that these changes are fully
covered by the total uncertainty of the luminosity
measurement of $2.3\%$.
The measurements of beauty production at threshold
\cite{Aaron:2012cj} and of elastic and proton-dissociative $J/\psi$
production \cite{Alexa:2013xxa} have systematic uncertainties which
are much larger than the correction discussed above and are not
updated.
In contrast, the measurements of inclusive neutral and
charged current cross sections \cite{Aaron:2012qi} reach a level of
precision where the $1.8\%$ correction may be of relevance.
Furthermore, the combined data tables 29-32, 45-48 and 51-52 in
\cite{Aaron:2012qi}, cannot be derived using a simple scale factor,
because other datasets, not affected by the problems discussed above,
are included in the averaging procedure.
The corrected data tables of \cite{Aaron:2012qi} are available \cite{download}.

We would like to thank Prof.~H.~Spiesberger for useful discussions on the
running of the fine-structure constant.

\begin{table}[h]
\begin{center}
\begin{tabular}{l||c|c||c|c}
\hline
 & \multicolumn{2}{c||}{no $\vert\vec{P}_T^{\text{miss}}\vert$ cut} &
 \multicolumn{2}{c}{$\vert\vec{P}_T^{\text{miss}}\vert<0.3\,\text{GeV}$} \\
 & \multicolumn{1}{c}{in ref. \cite{Aaron:2012kn}} & this analysis &
   \multicolumn{1}{c}{in ref. \cite{Aaron:2012kn}} & this analysis \\
\hline
quasi-elastic QEDC      & $6.84\%$ & $6.93\%$ & $2.93\%$ & $2.96\%$ \\
inelastic QEDC          & $7.02\%$ & $7.15\%$ & $1.51\%$ & $1.52\%$ \\
elastic DVCS            & $2.10\%$ & $2.06\%$ & $1.26\%$ & $1.24\%$ \\
quasi-elastic DVCS      & $0.55\%$ & $0.54\%$ & $0.16\%$ & $0.15\%$ \\
$ep \to ep\, e^{+}e^{-}$  & $1.15\%$ & $1.12\%$ & $1.31\%$ & $1.28\%$ \\
diffractive DIS         & $2.78\%$ & $2.72\%$ & $0.53\%$ & $0.52\%$ \\
non-diffractive DIS     & $0.02\%$ & $0.02\%$ & $0.01\%$ & $0.01\%$ \\
diffractive $\rho^0$    & $2.05\%$ & $2.00\%$ & $0.15\%$ & $0.15\%$ \\
diffractive $\omega$    & $0.43\%$ & $0.42\%$ & $0.03\%$ & $0.03\%$ \\
diffractive $\phi$      & $0.29\%$ & $0.28\%$ & $0.02\%$ & $0.02\%$ \\
diffractive $J/\psi$    & $0.20\%$ & $0.20\%$ & $0.05\%$ & $0.05\%$ \\
diffractive $\psi^{\prime}$& $0.17\%$ & $0.17\%$ & $0.08\%$ & $0.08\%$ \\
diffractive $\Upsilon$  & $0.02\%$ & $0.02\%$ & $0.01\%$ & $0.01\%$ \\
\hline
\end{tabular}
\caption{\label{table:background}Updated table of background fractions
  obtained from the reanalysis of \cite{Aaron:2012kn}.}
\end{center}
\end{table}
%

\begin{flushleft}

\end{flushleft}

\newcommand{\bhluminosity}{\ensuremath{338.9\pm10.2\,\text{pb}^{-1}}}
\newcommand{\totalsystheor}{\ensuremath{1.1\%}} 
\newcommand{\totalsysbgr}{\ensuremath{1.2\%}} 
\newcommand{\totalsysexp}{\ensuremath{1.4\%}} 
\newcommand{\totalerrstat}{\ensuremath{0.8\%}} 
\newcommand{\totalerrsys}{\ensuremath{2.1\%}} 
\newcommand{\totalerr}{\ensuremath{2.3\%}} 
\newcommand{\triggererror}{\ensuremath{0.2\%}}
\newcommand{\isrerror}{\ensuremath{1.0\%}}
\newcommand{\tpeerror}{\ensuremath{0.1\%}}
\newcommand{\gegmerror}{\ensuremath{0.2\%}}
\newcommand{\mcstaterror}{\ensuremath{0.3\%}}
\newcommand{\escaleerror}{\ensuremath{0.6\%}}
\newcommand{\ereserror}{\ensuremath{1.1\%}}
\newcommand{\posreserror}{\ensuremath{0.3\%}}
\newcommand{\cipefferror}{\ensuremath{0.2\%}}
\newcommand{\cipconverror}{\ensuremath{0.3\%}}
\newcommand{\alignmenterror}{\ensuremath{0.4\%}}
\newcommand{\zvtxerror}{\ensuremath{0.1\%}}
\newcommand{\ctdefferror}{\ensuremath{0.03\%}}
\newcommand{\clustererror}{\ensuremath{0.04\%}}
\newcommand{\larvetoerror}{\ensuremath{0.05\%}}
\newlength{\figwidth}
\newcommand{\textchanged}{\color{black}}
\setlength{\figwidth}{0.9\textwidth}

\def\Journal#1#2#3#4{{#1}~{\bf #2} (#3) #4}
\def\NCA{\em Nuovo Cimento}
\def\NIM{\em Nucl. Instrum. Methods}
\def\NIMA{{\em Nucl. Instrum. Methods} {\bf A}}
\def\NPB{{\em Nucl. Phys.}   {\bf B}}
\def\PLB{{\em Phys. Lett.}   {\bf B}}
\def\PRL{\em Phys. Rev. Lett.}
\def\PRD{{\em Phys. Rev.}    {\bf D}}
\def\ZPC{{\em Z. Phys.}      {\bf C}}
\def\EJC{{\em Eur. Phys. J.} {\bf C}}
\def\CPC{\em Comp. Phys. Commun.}

\begin{titlepage}

\noindent
\begin{flushleft}
{\tt DESY 12-062    \hfill    ISSN 0418-9833} \\
{\tt April 2012}                  \\
\end{flushleft}

\noindent

\vspace{2cm}
\begin{center}
\begin{Large}
{\bf Determination of the Integrated Luminosity at HERA using Elastic
  QED Compton Events}

\vspace{2cm}

H1 Collaboration

\end{Large}
\end{center}

\vspace{2cm}

\begin{abstract}
A measurement of the integrated luminosity at the $ep$ collider HERA is
presented, exploiting the elastic QED Compton process $ep\to e\gamma
p$.
The electron and the photon are detected in the backward
calorimeter of the H1 experiment.
The integrated luminosity of the data recorded in 2003 to 2007 is
determined with a precision of \totalerr{}.
The measurement is found to be compatible with the
corresponding result obtained using  the Bethe-Heitler process.

\end{abstract}

\vspace{1.5cm}

\begin{center}
Published in \EJC{} 72 (2012) 2163
\end{center}

\end{titlepage}

%
%
%
%
\noindent
F.D.~Aaron$^{5,48}$,           
C.~Alexa$^{5}$,                
V.~Andreev$^{25}$,             
S.~Backovic$^{30}$,            
A.~Baghdasaryan$^{38}$,        
S.~Baghdasaryan$^{38}$,        
E.~Barrelet$^{29}$,            
W.~Bartel$^{11}$,              
K.~Begzsuren$^{35}$,           
A.~Belousov$^{25}$
% ...


\bigskip{\it
\noindent
% ...
$ ^{5}$ National Institute for Physics and Nuclear Engineering (NIPNE) ,
          Bucharest, Romania$^{ m}$ \\
% ...
$ ^{11}$ DESY, Hamburg, Germany \\
% ...
$ ^{25}$ Lebedev Physical Institute, Moscow, Russia \\
% ...
$ ^{29}$ LPNHE, Universit\'e Pierre et Marie Curie Paris 6,
          Universit\'e Denis Diderot Paris 7, CNRS/IN2P3, Paris, France \\
 $ ^{30}$ Faculty of Science, University of Montenegro,
          Podgorica, Montenegro$^{ n}$ \\
% ...
$ ^{35}$ Institute of Physics and Technology of the Mongolian
          Academy of Sciences, Ulaanbaatar, Mongolia \\
% ...
$ ^{38}$ Yerevan Physics Institute, Yerevan, Armenia \\
% ...
$ ^{48}$ Also at Faculty of Physics, University of Bucharest,
          Bucharest, Romania \\
% ...
$ ^m$ Supported by the Romanian National Authority for Scientific Research
      under the contract PN 09370101 \\
 $ ^n$ Partially Supported by Ministry of Science of Montenegro,
      no. 05-1/3-3352 \\
}

\newpage


\section{Introduction}

For particle collider experiments, the precise knowledge of the
luminosity is essential for any type of cross section
measurement.
%
The instantaneous luminosity is defined as 

\begin{equation}
L=\frac{fnN_{1}N_{2}}{A},
\end{equation}
where $f$ is the revolution frequency for the two colliding particles $p_1$ and $p_2$,
$n$ is the number of colliding bunches
per revolution, and $N_1$ ($N_2)$ is the number of particles of type $p_1$
($p_2$) per bunch. The effective cross section of the beams is $A$.
The time-integrated luminosity ${\cal L}$ relates the cross section
$\sigma_{p_1p_2\to X}$ of the reaction $p_1p_2\to X$ to the number of events 
$N_{p_1p_2\to X}$ expected in the time interval $T$ by
\begin{equation}
\label{eqn:intlumi}
{\cal L}=\int_T Ldt = \frac{N_{p_1p_2\to X}}{\sigma_{p_1p_2\to
    X}}.
\end{equation}
Since it is difficult to monitor all beam parameters with a per cent
level precision, in particular those defining the effective beam cross
section $A$, the integrated luminosity is often determined by counting
the number of observed events for a specific reaction $p_1p_2\to X$
with a well-known cross section.

At HERA, the colliding beams are protons and electrons\footnote{
In this paper the term
``electron'' is used generically to refer to both electrons and
positrons.}.
For the data taking period studied in this analysis, the proton beam
energy is $E^0_p=920\,\mathrm{GeV}$ and the electron beam energy is
$E^0_e=27.6\,\text{GeV}$.
The reaction used to determine the integrated luminosity is the
production of a radiative photon in elastic $ep$ scattering, $ep\to
e\gamma p$.
Depending on the phase space considered, this process is referred to
as Bethe-Heitler (BH) scattering or QED
Compton (QEDC) scattering.
In the BH process\cite{Bethe:1934za}, both the electron and the photon are
emitted almost collinearly to the incident electron.
The corresponding cross section is very large, ${\cal
  O}(100\,\text{mb})$. Dedicated small angle detectors are used to record BH
events.
In contrast, for QEDC scattering\cite{qedctheory}, the particles have a sizable
transverse momentum with respect to the incident electron
and can be detected in the main detector.
The momentum transfer squared at the proton vertex, $t$, is generally small.
At very small momentum transfer $\vert t\vert \ll 1\,\text{GeV}^2$, elastic
scattering dominates.
At $\vert t\vert \gtrsim 1\,\text{GeV}^2$, inelastic processes are
relevant and the reaction is sensitive to the proton
structure.
In addition, there are quasi-elastic contributions to the cross
section, where the outgoing proton forms an excited state, like
$\Delta^{+}$ or $N^{\star}$, which then decays to a low mass hadronic system.
Within the phase space considered in this analysis, the elastic QEDC cross
section is ${\cal O}(50\,\text{pb})$. 

At HERA, the integrated luminosity is usually measured in the 
BH process, using dedicated detectors located at small angles.
The advantage of this process is its very large cross section, thus 
negligible statistical uncertainties are achieved for small amounts of
integrated luminosity.
However, there are various sources of possibly large systematic uncertainty.
For example, there may be inevitable acceptance limitations for the small angle
detectors, caused by elements of the beam transport system which 
separates the BH photons and electrons from the circulating proton and
electron beams.
The acceptances of the photon and electron detectors may exhibit
complex spatial structures and can vary in time as well.
Another complication originates from synchrotron radiation emitted by
the electron beam as it passes the focusing magnets surrounding the
interaction region.
Furthermore $ep$ collisions can happen also outside the nominal
interaction region. These contribute to the BH measurement of the
integrated luminosity, but must be corrected for when analysing cross
sections with the H1 main detector, which has a more limited acceptance
as a function of the collision point position.

In this paper, a determination of the integrated luminosity is
presented, based on the elastic QEDC process, which is measured in the H1 main
detector.
This method is insensitive to details of the beam optics. However, the
smallness of the cross section leads to limited statistical precision, 
thus time-dependencies
can not be resolved with high resolution by counting elastic QEDC events alone.
Auxiliary measurements of other reactions with higher cross section may
be used to monitor time-dependencies.

Comparisons of BH and QEDC measurements at HERA have been performed
previously \cite{Ahmed:1995cf}.
The inelastic QEDC process also has been measured at large $\vert t \vert$
\cite{Aktas:2004ek}.
The data available for the elastic QEDC analysis described in the following
were recorded with the H1 detector in the years 2003 to 2007.

\section{H1 Detector}

\begin{figure}[htbp]
  \begin{center}
\includegraphics[width=\figwidth]{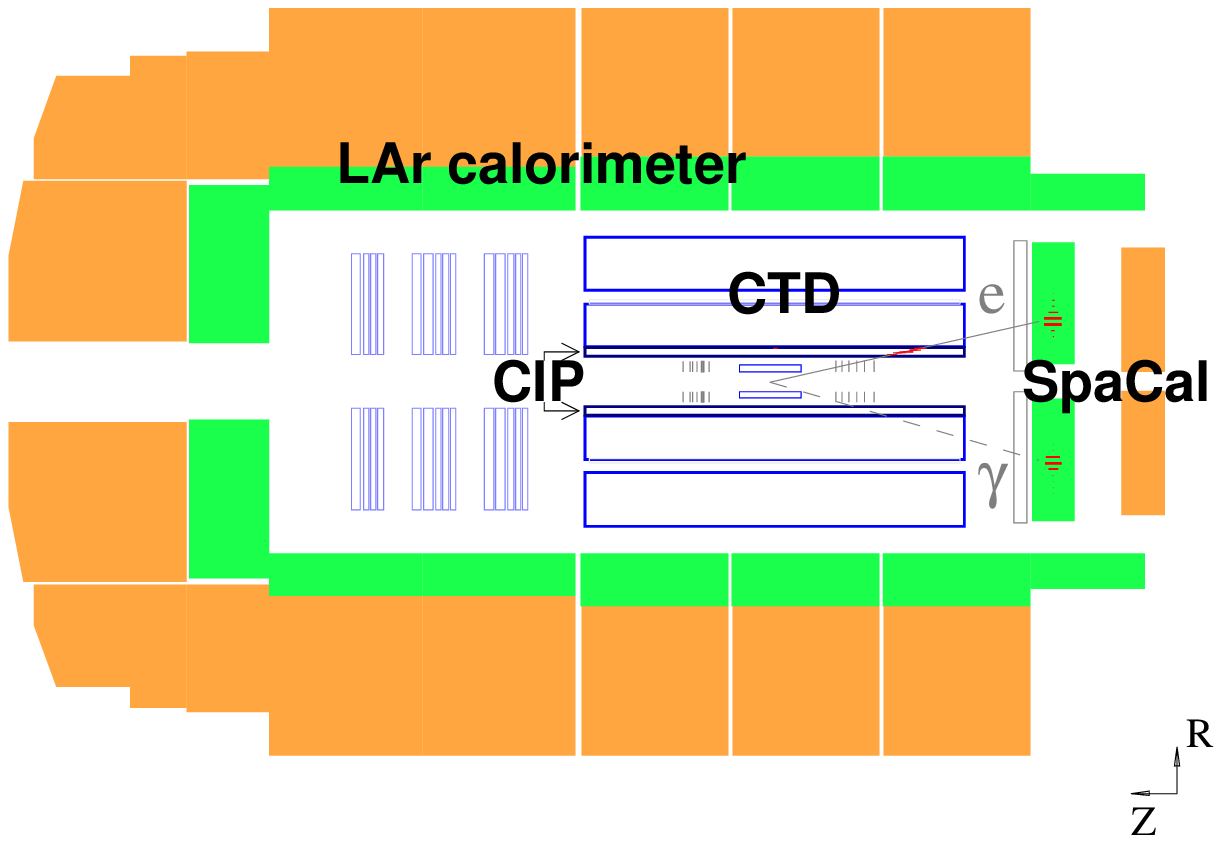}
\caption{\label{fig:event}Elastic QEDC candidate event observed in
  the H1 detector. The H1 detector components most relevant to this analysis
  are indicated. The approximate electron and photon candidate
  trajectories are shown.}
\end{center}
\end{figure}

In the following, only those components of the H1 detector are briefly
introduced which are essential for the present analysis.
A detailed description of the whole detector in its original configuration can
be found elsewhere \cite{H1detector}.
Components which were part of later upgrades are referred
to here separately.
%
%
The origin of the H1 coordinate system is the nominal
$ep$ interaction point.
The direction of the proton beam defines the positive $z$--axis
(forward direction).
Transverse momenta are measured in the $xy$ plane.
Polar~($\theta$) and~azimuthal~($\varphi$) angles are measured with
respect to this reference system.
The pseudo-rapidity is defined as $\eta= -\ln{\tan (\theta/2)}$.
A schematic view of the H1 detector with signals from
an elastic QEDC candidate event is shown in figure \ref{fig:event}.

In the backward region $-4.0 < \eta < -1.4$, a lead-scintillating
fibre calorimeter\cite{spacal} (SpaCal) is used for the identification and
measurement of both the scattered electron and the scattered photon.
The energy resolution for electromagnetic showers is $\sigma(E)/E
\simeq 7.1\%/\sqrt{E/{\rm GeV}} \oplus 1\%$ \cite{spacaltest}.
The electromagnetic section of the SpaCal is read out in cells of size
$4\times 4\,\text{cm}$ in the $xy$ plane, where the 
Moli\`{e}re radius is $2.5\,\text{cm}$.
The $xy$ position of a shower is reconstructed as a weighted
mean of the cell centres, the weights taken proportional to the
logarithm of the cell energies \cite{Glazov:2010zza}.
After applying $xy$ dependent corrections, the position resolution is
about $3.5\,\text{mm}$ for an electromagnetic shower in
the energy range relevant to this analysis. 

The liquid argon (LAr) calorimeter covers the range $-1.5 < \eta <
3.4$.
Its electromagnetic (hadronic) section is equipped with absorbers made
of lead (steel) plates.
An energy resolution 
of $\sigma(E)/E \simeq 11\%/\sqrt{E/{\rm GeV}}$ 
for electromagnetic showers and 
of $\sigma(E)/E \simeq 50\%/\sqrt{E/{\rm GeV}}$ 
for hadronic showers is obtained from
test beam measurements.

The central region of the detector is equipped with a set of tracking
detectors (CTD).
There are the two concentric central jet chambers (CJC), interleaved
by a $z$ chamber, and the central silicon tracker (CST)~\cite{Pitzl:2000wz}.
The CTD measures the momenta of charged particles in the angular range
$20^{\circ} < \theta < 160^{\circ}$.
The central inner proportional chambers (CIP)~\cite{Becker:2007ms} are
located between the inner CJC and the CST.
The five CIP chambers have a radial spacing of
$9\,\text{mm}$, where the innermost layer is located at a radius of
$15.7\,\text{cm}$.
In $\varphi$ there is a $16$-fold segmentation, whereas in $z$ the segments
have variable size, ranging from $1.8\,\text{cm}$ in the innermost
layer to $2.3\,\text{cm}$ in the outermost layer.
%
The CIP has an angular acceptance in the range
$10^{\circ}<\theta<170^{\circ}$.
In the backward region, the tracking is complemented by the backward
proportional chamber (BPC), located directly in front of the SpaCal.

The calorimeters and tracking detectors are located inside
a large superconducting solenoid, providing a uniform field of
$1.16\,\text{T}$ strength.
The return yoke of the solenoid is
instrumented and serves as a muon detector.
Upstream and downstream of the interaction point there are systems of
scintillators (VETO), providing time-of-flight information.
Timing signals from the VETO and the SpaCal were used during data
taking to reject particles originating from non-$ep$ interactions of
the proton beam in the HERA tunnel.
The luminosity system for measuring the Bethe-Heitler process 
consists of an electron tagger located at  $z=-5.4\,\text{m}$ and a photon
calorimeter located at $z = -103 \ {\rm m}$.

\section{Signal and Background processes}

Monte Carlo event generators (MC) are used to predict event yields of
signal and background processes.
A GEANT3 \cite{geant} simulation of the H1 detector is
performed for each generated event, where also the relevant time-dependencies
such as changes to the detector setup and varying beam conditions are taken
into account.
Electromagnetic showers are simulated using a shower
library\cite{Glazov:2010zza}.
After detector simulation, the events are passed through the same
reconstruction algorithms as were used for the data.

The QEDC signal is simulated using the COMPTON22 event generator
\cite{compton22}.
This generator produces elastic, quasi-elastic and inelastic events.
The elastic QEDC events are taken as signal, since their cross
section only depends on QED theory and on the proton elastic form factors,
thus having small uncertainties.
{\textchanged Details are discussed in Section \ref{text:theoryerror}.}
The quasi-elastic events are treated as background and suppressed in the analysis,
because their cross section depends on less precisely known
parameters such as probabilities to produce excited nucleons.
Similarly, the inelastic events are treated as background, because
their cross section depends, for example, on {\textchanged parameterisations} of the proton 
structure functions at very low momentum transfer.
In COMPTON22, the fragmentation of quasi-elastic events is modelled
using the SOPHIA package \cite{Mucke:1999yb}, whereas for inelastic
events string fragmentation as implemented in PYTHIA
\cite{Sjostrand:2000wi} is used.
For the elastic QEDC signal, final state radiation from the electron
has been included in the COMPTON22 event generator using the relevant PYTHIA
routines.

An important source of background is electron-positron pair production,
$ep\to ep\,e^{-}e^{+}$, simulated using the GRAPE
event generator \cite{Abe:2000cv}. 
Other background events originate mainly from various diffractive
processes, namely deeply virtual Compton scattering (DVCS), diffractive
vector meson (VM) production and non-resonant diffraction.
DVCS is modelled using the MILOU event generator \cite{milou}.
Diffractive VM production is simulated using the DIFFVM
event generator \cite{diffvm}, where the production of $\rho^0$, $\omega$,
$\phi$, $J/\psi$, $\psi^{\prime}$ and $\Upsilon$ mesons is considered.
For $\rho^0$ production, DIFFVM is modified such that decays to
$\pi^0\gamma$ and $\eta^0\gamma$ are included.
Non-resonant diffraction is simulated using the RAPGAP event generator
\cite{rapgap}.
Background from non-diffractive deep-inelastic scattering (DIS) is simulated
using DJANGO \cite{Schuler:1991yg} and is found to be negligible.

\section{Event selection}

During data taking, events with electrons in the SpaCal were recorded 
if certain trigger criteria were fulfilled.
The trigger required  the presence of at least one localised energy
deposit in the SpaCal calorimeter with energy $E>6\,\text{GeV}$. In
addition, there were veto conditions to reject beam related background
not originating from $ep$ collisions. Data periods where the SpaCal
trigger was not fully efficient
are removed from the analysis,
corresponding to about $6\%$ of the total H1 data sample recorded in
2003 to 2007.

The elastic QEDC events are selected offline by requiring two
clusters in the electromagnetic section of the SpaCal.
A summary of the selection criteria is given in table \ref{tab:cuts}.
The transverse sizes of the SpaCal clusters are restricted to
$R_{\log{}}<6\,\text{cm}$, where $R_{\log{}}$ is calculated from the SpaCal
cell centres using logarithmic energy weighting \cite{Glazov:2010zza}.
The cluster energies are required to be larger than $2.2\,\text{GeV}$.
In the range $30\le R<72\,\text{cm}$ of the radial distance from the
beam, $R$, exactly two such clusters are
required, whereas for $20\le R< 30\,\text{cm}$ no cluster is allowed.
The restriction in the number of clusters removes 
background and events with hard radiative photons.
The restriction $R\ge 30\,\text{cm}$ on the two clusters ensures that
the particles are within the CIP acceptance.

Electron trajectories are reconstructed using the SpaCal cluster
position together with position information from the CIP chambers.
Hits in the CIP chambers are considered if they match the SpaCal cluster
in $\varphi$.
Adjacent hits are merged to CIP clusters in $\varphi$ and $z$.
This merging is done separately for each layer.
A straight line fit of the CIP clusters and the SpaCal cluster in the
$rz$ plane is performed, where outliers are rejected.
The coordinate $r$ is the radial distance from the $z$ axis, measured along the
azimuthal direction given by the SpaCal cluster.
After outlier rejection, there are up to five accepted CIP clusters,
corresponding to the five CIP layers.
At least two accepted CIP clusters are required.
Next, the centre-of-gravity of the CIP clusters in the
$rz$ plane is calculated.
Finally, the CIP centre-of-gravity in $rz$ , together with
the SpaCal energy and the SpaCal position are used to reconstruct a helix
trajectory in three dimensions, pointing back to the origin of the
interaction.
In the determination of this helix, the beam
spot and beam tilt are also used\footnote{The beam spot is defined as
the average $x$ and $y$ position of interactions which take place at
$z=0$. The beam tilt is a slope correction for interactions at
$z\ne 0$. These parameters were monitored regularly in short time
intervals using the CTD.}.
The direction of bending in the magnetic field
is chosen assuming that the particle charge is equal to the
charge of the beam lepton.
The algorithm finally returns the origin of the interaction
(CIP vertex) and the momentum vector at the CIP vertex.

The electron and the photon are then identified, making use of the
helix fit results. If there is no CIP vertex, the event is rejected.
If there is only one SpaCal cluster linked to a CIP vertex, that cluster is
taken as electron while the other cluster is taken as photon.
The photon momentum vector is calculated from the photon 
cluster energy and a straight line trajectory pointing from the 
electron CIP vertex to the photon cluster position.
If both SpaCal clusters are linked to CIP vertices, it is assumed that
the photon has converted into an electron-positron pair while passing
the material in front of the CIP detector.
For that case, two hypotheses are checked in the reconstruction.
First, one of the clusters is taken as the electron, and the photon momentum
is calculated using the energy and position of the other cluster as
described above.
The difference in azimuth between the electron and photon candidate 
momenta, $\Delta\varphi_1$, is determined.
Next, the particle hypotheses are interchanged 
and the corresponding difference in azimuth, $\Delta\varphi_2$, is calculated.
The particle assignment is done according to the hypothesis yielding
an azimuthal opening angle closer to $180^{\circ}$.
In the simulation, the mis-identification probability is
$0.3\%$ ($16\%$) if one (both) SpaCal clusters are linked to CIP vertices.
\begin{table}[t]
\begin{center}
\begin{tabular}{l}
\hline
Electromagnetic SpaCal clusters with transverse sizes $R_{\log{}}<6\,\text{cm}$ and $E>2.2\,\text{GeV}$ \\
Exactly two clusters with radial distance to the beam axis $30\le R<72\,\text{cm}$ \\
No additional cluster with $20\le R<30\,\text{cm}$ with energy $E>2.2\,\text{GeV}$ \\
At least one of the two clusters with CIP vertex \\
Electron and photon identification \\
Energies $\min(E_e,E_\gamma)>7\,\mathrm{GeV}$ and 
$\max(E_e,E_\gamma)>10\,\mathrm{GeV}$ \\
Vertex position $\vert z_{\text{vtx}}\vert<35\,\text{cm}$ \\
Polar angles $155.9^{\circ}<\theta_e,\theta_\gamma<169.5^{\circ}$ \\
Difference in azimuth $170^{\circ}<\Delta\varphi<190^{\circ}$ \\
Transverse momentum balance $\vert \vec{P}_T^{\text{miss}}\vert <0.3\,\text{GeV}$ \\
Energy in LAr at $\theta<10^{\circ}$ is below $0.5\,\text{GeV}$ \\
No additional CTD track \\
\hline
\end{tabular}
\caption{\label{tab:cuts}
Summary of selection criteria.}
\end{center}
\end{table}

Once the electron and photon are identified, the $z$
position of the electron CIP vertex, $z_{\text{vtx}}$, is verified.
Only events with $\vert z_{\text{vtx}}\vert<35\,\text{cm}$ are
selected in the analysis.
To further suppress background contributions the following cuts are applied:
energy of the most (least) energetic particle greater than $10$~($7$)~$\text{GeV}$,
polar angles $\theta_e,\theta_\gamma$ within $155.9^{\circ}$ and $169.5^{\circ}$,
difference in azimuth between $170^{\circ}$ and $190^{\circ}$, 
modulus of the transverse component of the missing momentum
smaller than $0.3\,\text{GeV}$. The missing
momentum vector is calculated from the reconstructed electron and
photon four-momenta.

The inelastic background sources are further suppressed by using
conditions on additional activity in the detector.
Events are rejected if
the energy in the forward part of the LAr calorimeter,
with polar angle $\theta<10^{\circ}$, exceeds $0.5\,\text{GeV}$, or if there
are 
CTD tracks 
which can not be
attributed to either the electron or the photon.
The total number of elastic QEDC candidate events is $14277$, after
correcting for trigger efficiency effects.

The efficiency of the reconstruction is determined for the following 
phase space defined for the generated elastic QEDC signal:
the polar angular range of the generated electron and photon is restricted to
$155.9^{\circ}<\theta<169.5^{\circ}$, the maximum fraction of incident electron
energy carried away by initial state radiation is $0.35$, and the
momentum transfer squared at the proton vertex is limited to 
$\vert t\vert<0.09\,\text{GeV}^2$. For this generated phase space, the cross
section is $\sigma_{\text{gen}}=54.8\,\text{pb}$.
A total fraction of $f_{\text{vtx}}=2.5\%$ of the generated events have a simulated vertex position
along the $z$ coordinate, $z_{gen}$, outside $\pm 35\,\text{cm}$ around the nominal vertex position, in order to
be able to describe the observed longitudinal vertex distribution in
the H1 detector.
Within the region $\vert z_{gen}\vert<35\,\text{cm}$
the reconstruction efficiency is found to be $\epsilon_{\text{rec}}=64.7\%$. 
The sources of {\textchanged inefficiency} are investigated in the following.
Losses of $12.4\%$ originate from the cut in
$\vert\vec{P}_T^{\text{miss}}\vert$ due to the limited experimental 
resolution.
The requirement $30<R<72\,\text{cm}$ on the radial SpaCal cluster
position reduces the efficiency by $7.2\%$.
This loss is related to a geometrical effect, such that particles
originating from $z_{\text{gen}}>0$ ($z_{\text{gen}}<0$) and scattered
at polar angles near $155.9^{\circ}$ ($169.5^{\circ}$) are outside the
allowed range in $R$.
The track and calorimeter veto conditions contribute to the inefficiency by
$3.4\%$, dominated by the restriction in forward LAr energy and the veto on
additional SpaCal clusters.
The LAr condition is not fully efficient due to electronic noise and
overlap with non $ep$ background.
The inefficiency due to a third SpaCal cluster originates from events
with hard final state radiation.
Losses due to the other selection criteria are small ($1.8\%$ total).
Finally, $10.5\%$ are rejected by more than one selection
criterion, where combinations involving the
$\vert\vec{P}_T^{\text{miss}}\vert$ condition dominate.
\begin{table}[t]
\begin{center}
\begin{tabular}{l|c|c}
\hline
 & no $\vert\vec{P}_T^{\text{miss}}\vert$ cut &
 $\vert\vec{P}_T^{\text{miss}}\vert<0.3\,\text{GeV}$ \\
\hline
quasi-elastic QEDC       & $6.84\%$ & $2.93\%$ \\
inelastic QEDC           & $7.02\%$ & $1.51\%$ \\
elastic DVCS             & $2.10\%$ & $1.26\%$ \\
quasi-elastic DVCS       & $0.55\%$ & $0.16\%$ \\
$ep \to ep\, e^{+}e^{-}$  & $1.15\%$ & $1.31\%$ \\
diffractive DIS          & $2.78\%$ & $0.53\%$ \\
non-diffractive DIS      & $0.02\%$ & $0.01\%$ \\
diffractive $\rho^0$       & $2.05\%$ & $0.15\%$ \\
diffractive $\omega$     & $0.43\%$ & $0.03\%$ \\
diffractive $\phi$       & $0.29\%$ & $0.02\%$ \\
diffractive $J/\psi$     & $0.20\%$ & $0.05\%$ \\
diffractive $\psi^{\prime}$ & $0.17\%$ & $0.08\%$ \\
diffractive $\Upsilon$   & $0.02\%$ & $0.01\%$ \\
\hline
\end{tabular}
\caption{\label{tab:bgr}
  Background processes contributing to the elastic QEDC
  selection. The background fractions are given for both a selection with
  the $\vert\vec{P}_T^{\text{miss}}\vert$ cut released and for the
  final selection.
}
\end{center}
\end{table}

A detailed breakdown of the different background sources contributing to the elastic QEDC selection as defined in table \ref{tab:cuts} is given in  table~\ref{tab:bgr} without and with applying the $\vert\vec{P}_T^{\text{miss}}\vert$ cut.
More than half of the background originates from 
inelastic and quasi-elastic QEDC processes.
The $\vert\vec{P}_T^{\text{miss}}\vert$ cut significantly reduces the background by about a factor of three. 

For measuring the integrated luminosity ${\cal L}^{\text{QEDC}}$ of
collisions originating from the region $\vert z_{gen}\vert<35\,\text{cm}$, the
following relation is used 
\begin{equation}
{\cal L}^{\text{QEDC}} = (1-f_{\text{vtx}})\frac{N_{\text{event}}(1-f_{\text{bgr}})}{\sigma_{\text{vis}}},
\end{equation}
where $N_{\text{event}}$ is the number of QEDC candidate events observed in
the detector, $f_{\text{bgr}}$ is the background fraction predicted
by the MC simulation and $\sigma_{\text{vis}}=36.4\,\text{pb}$ is
the visible QEDC cross section. 
The main contribution to $\sigma_{\text{vis}}$ 
originates from genuine QEDC production in the phase space region of this analysis and with an interaction vertex within the accepted region 
$(1-f_{\text{vtx}})\sigma_{\text{gen}}\times\epsilon_{\text{rec}}=34.6\,\text{pb}$.
Additional contributions from 
events outside the defined phase space or with an interaction vertex beyond the defined limits amount to $1.8 \,\text{pb}$.

\begin{figure}[htbp]
  \begin{center}
\includegraphics[width=0.45\figwidth]{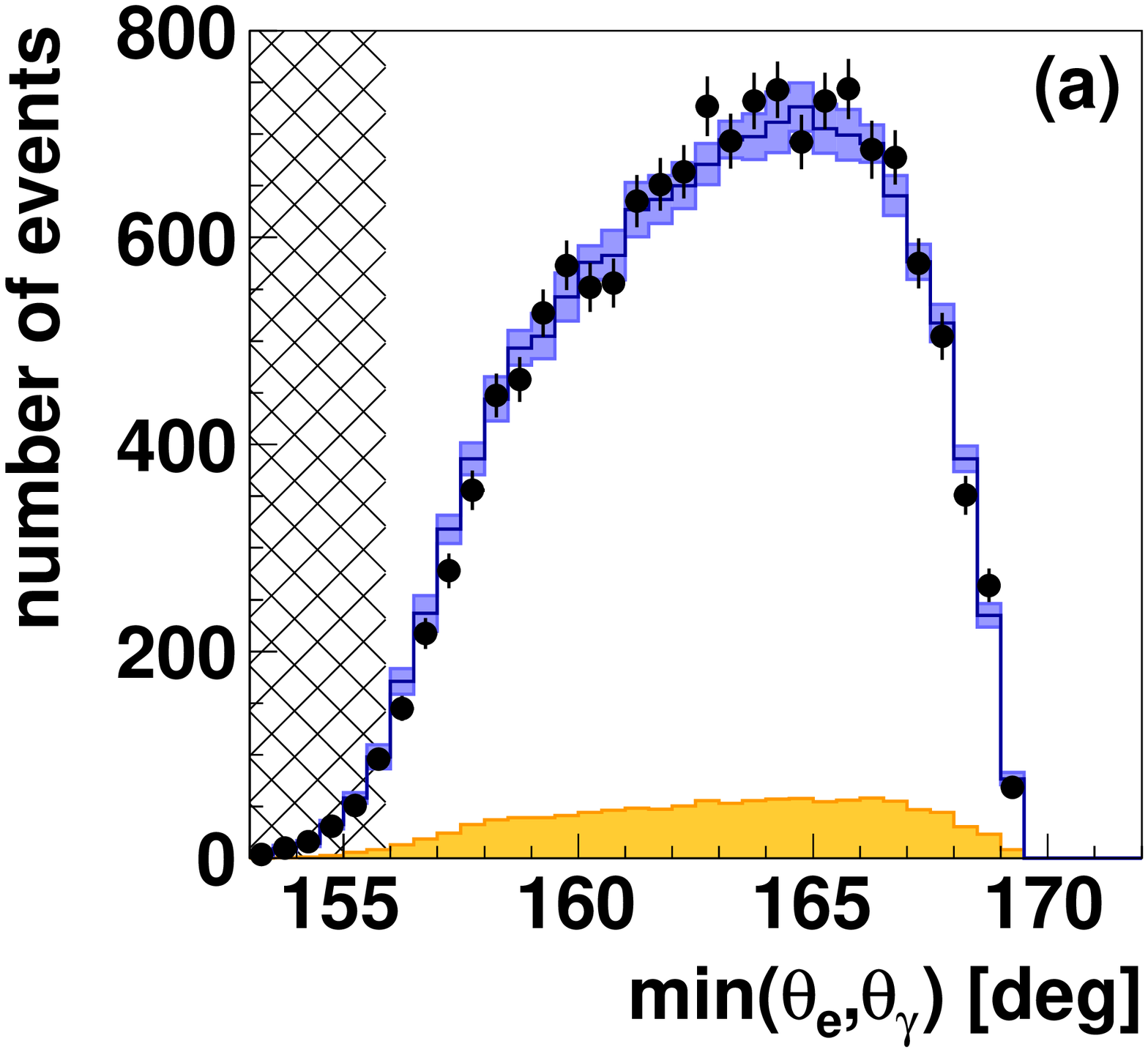}
\includegraphics[width=0.45\figwidth]{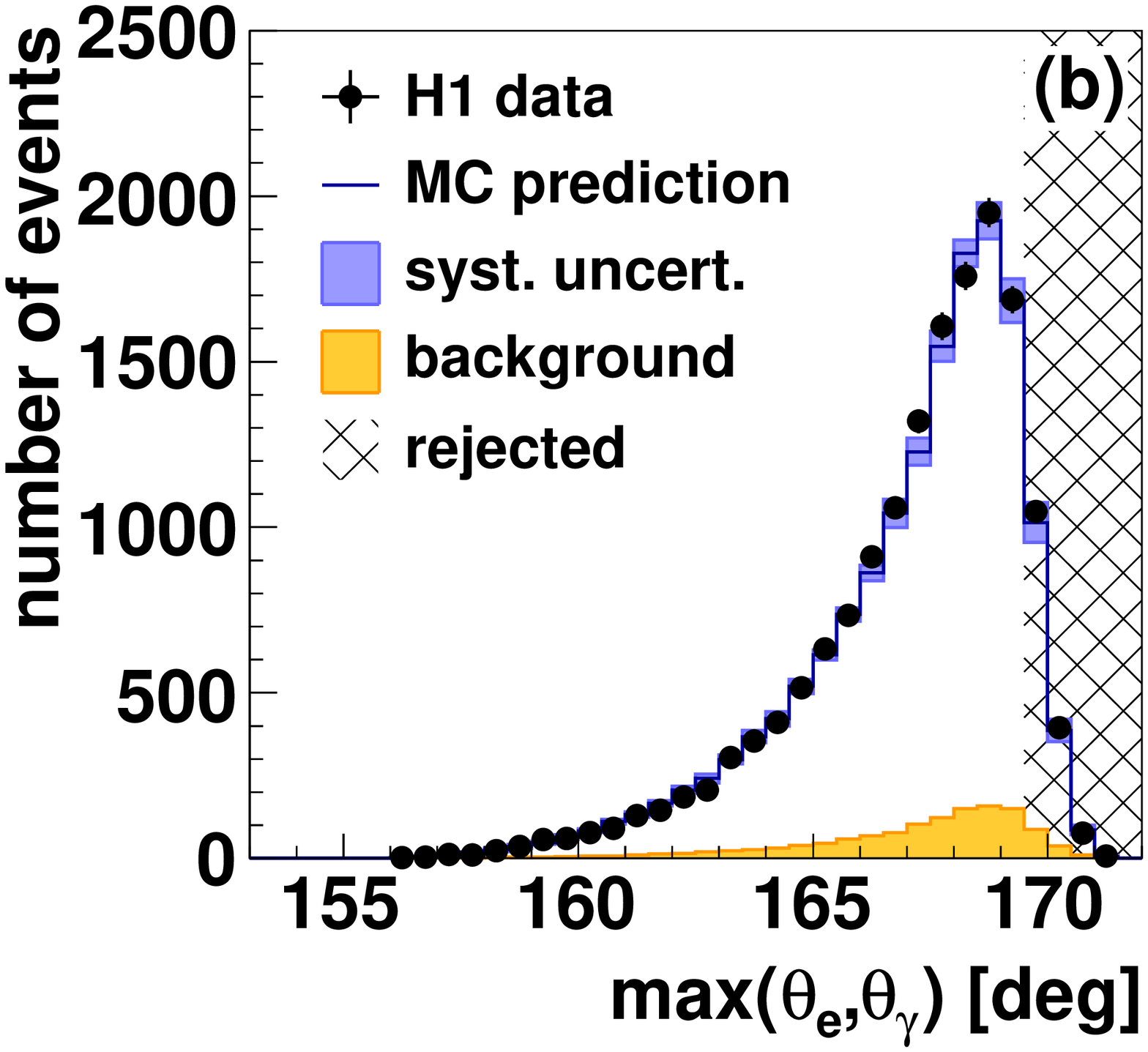}
\includegraphics[width=0.45\figwidth]{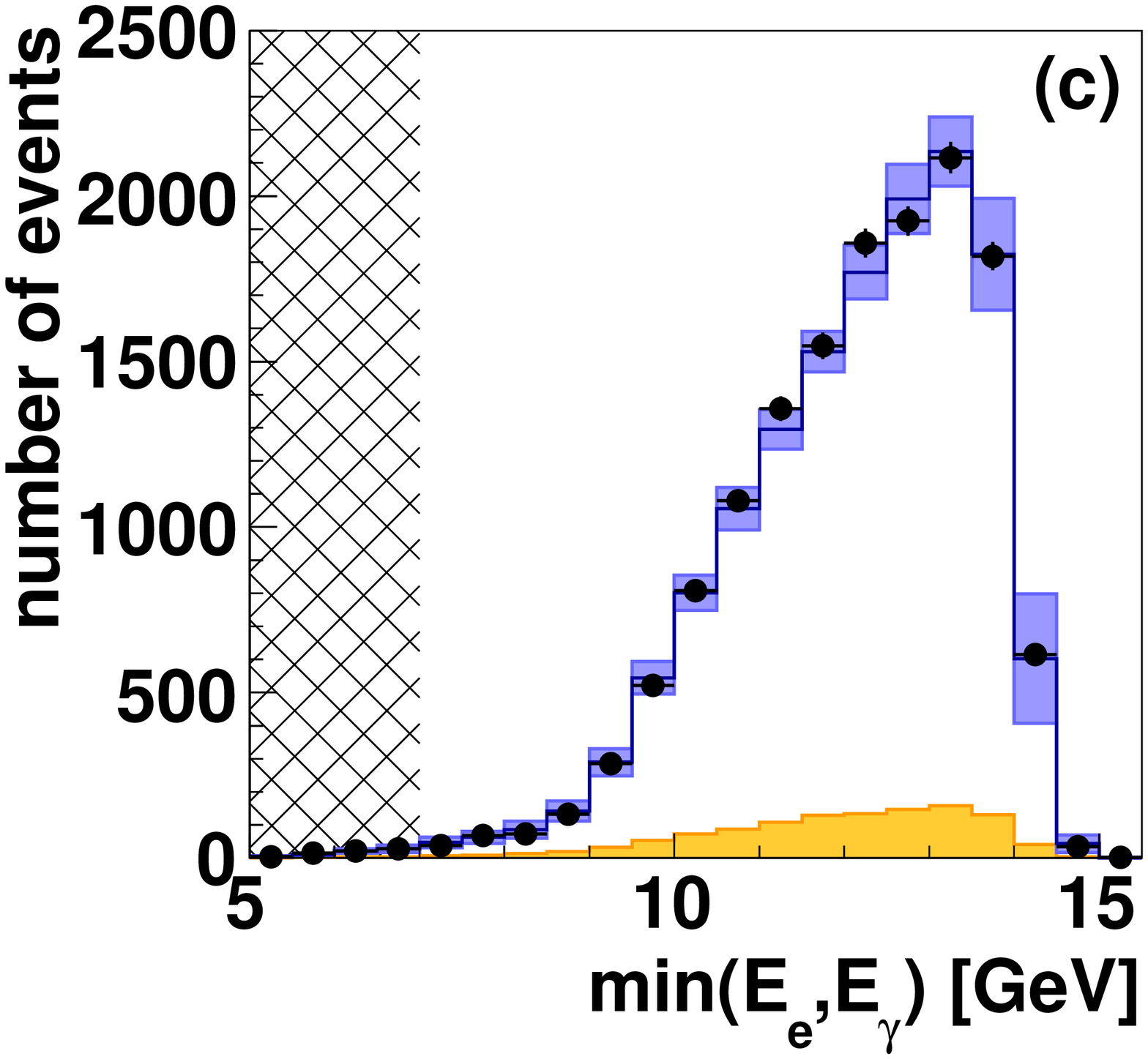}
\includegraphics[width=0.45\figwidth]{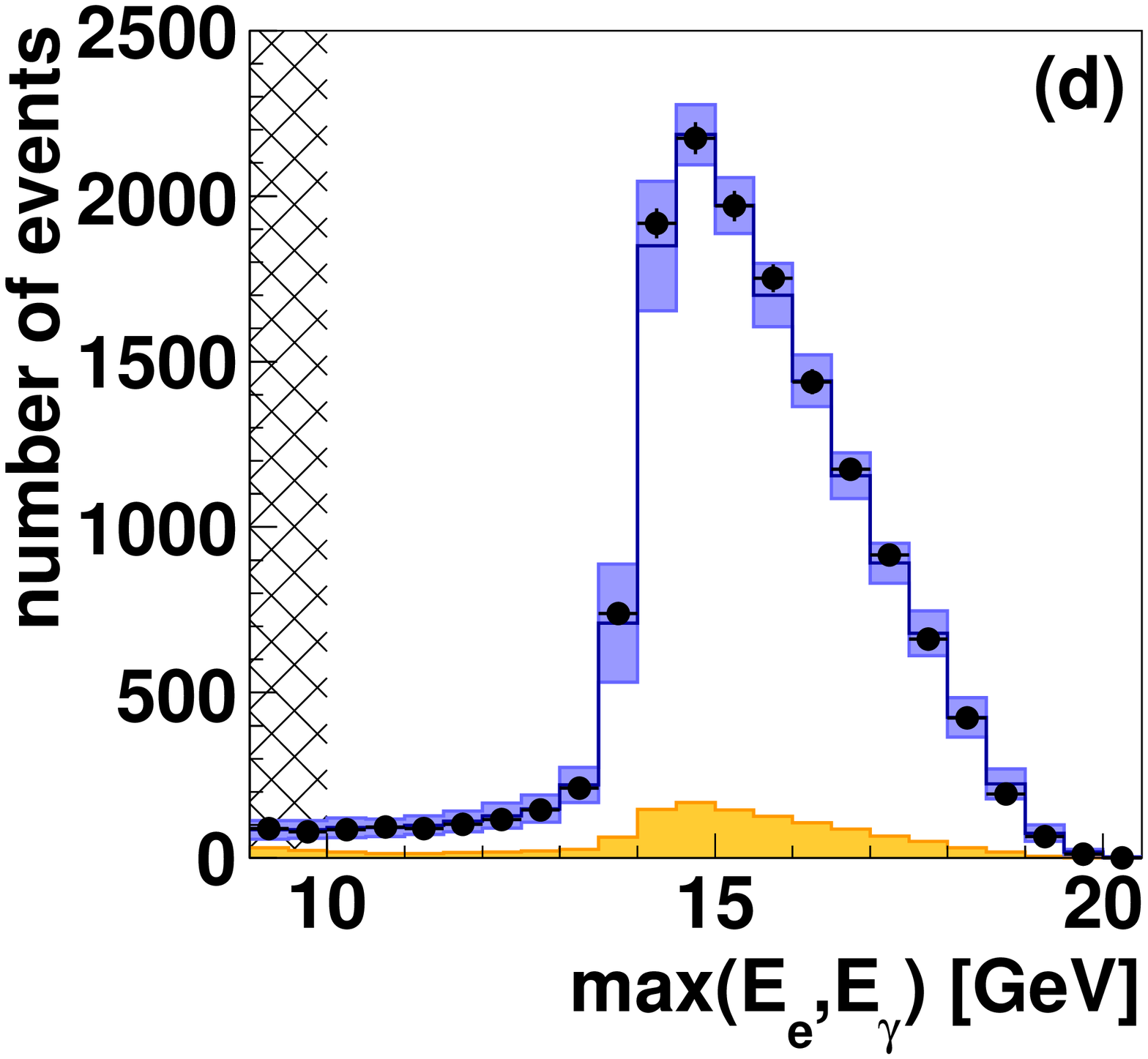}
\includegraphics[width=0.45\figwidth]{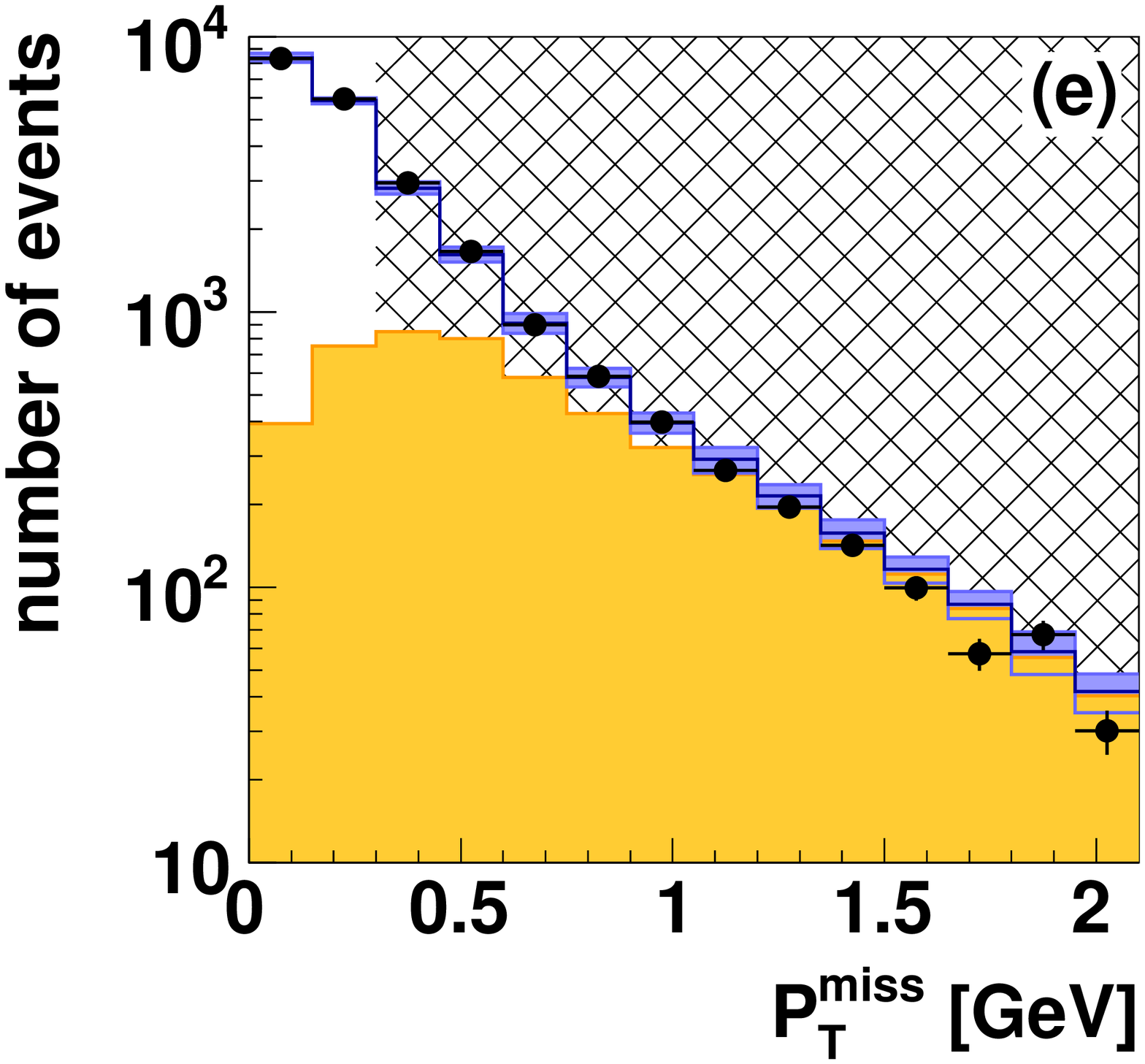}
\includegraphics[width=0.45\figwidth]{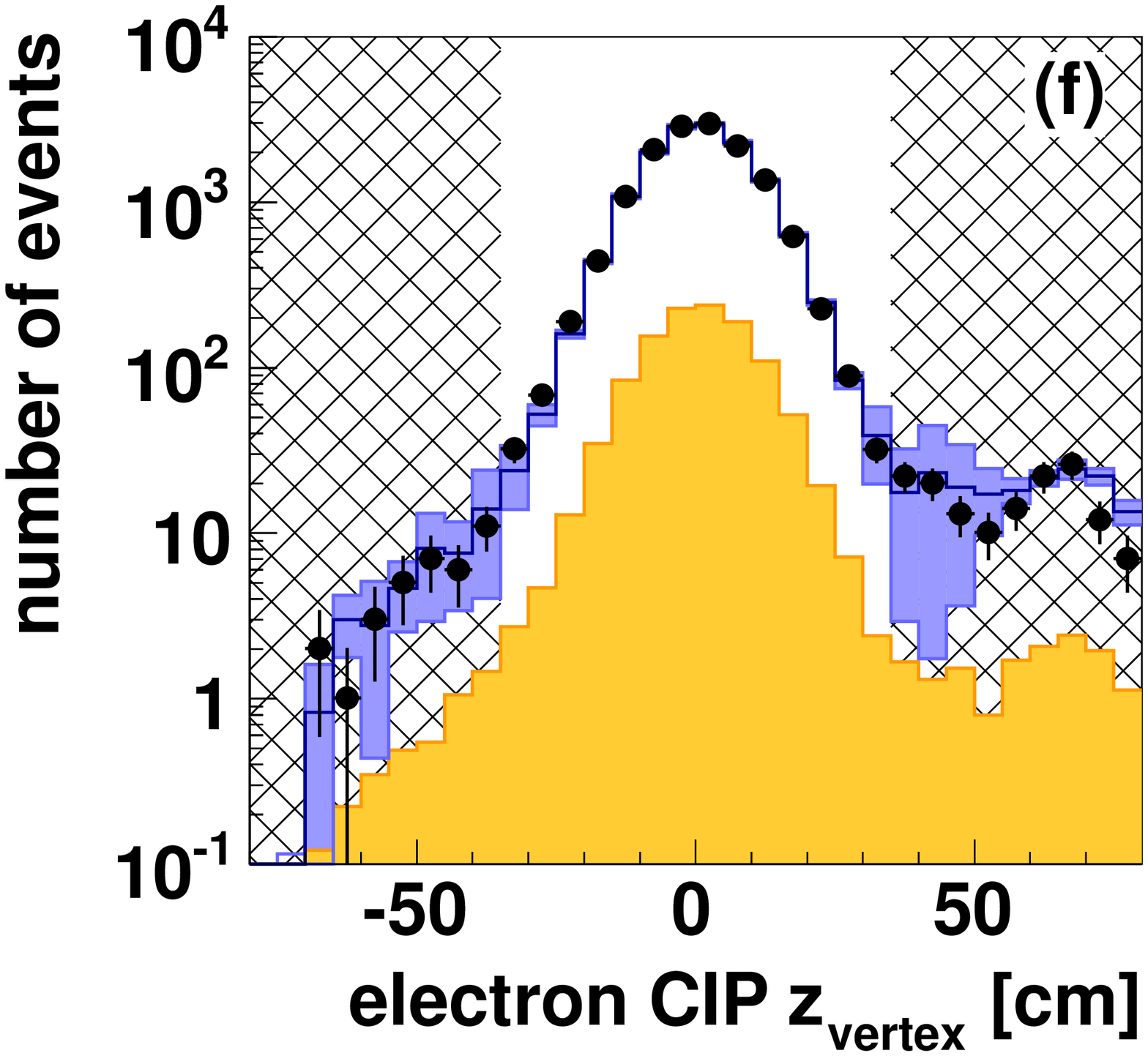}
  \caption{\label{fig:selection}
    Distributions of variables used to select elastic QEDC events:
    (a--e) kinematic quantities
    of the selected electron-photon pair and (f) the $z$ coordinate of the
    position of the interaction. The kinematic quantities are (a) the minimum
    polar angle, (b) the maximum polar angle, (c) the minimum energy, (d) the
    maximum energy and (e) the modulus of their total transverse momentum.
    The data are shown as black dots with the statistical uncertainties
    indicated as vertical bars. The simulation including background,
    normalised to
    the integrated luminosity determined in this analysis, is indicated as a solid
    line, with the systematic uncertainties attached as shaded area.
    Also shown is the contribution from background. The hatched areas
    are excluded by the selection criteria.}
\end{center}
\end{figure}

Distributions of variables used in the selection procedure are shown in figure
\ref{fig:selection}. Within uncertainties, the data are described by
the prediction.
Note that the prediction is normalised to the integrated luminosity ${\cal
  L}^{\text{QEDC}}$ as determined in the present analysis.
The $\vert \vec{P}_T^{\text{miss}}\vert$ distribution, figure \ref{fig:selection}e,
is of particular interest as it shows a clear separation of background
and signal.
The analysis cut of $0.3\,\text{GeV}$ is a compromise between
inevitable systematic uncertainties due to the limited detector
resolution, dominating at small  $\vert \vec{P}_T^{\text{miss}}\vert$,
and background contributions, increasing at large $\vert
\vec{P}_T^{\text{miss}}\vert$.

\section{Systematic uncertainties}

Systematic uncertainties on the elastic QEDC measurement may be categorised
as follows: experimental uncertainties,  background uncertainties and QEDC
theory uncertainties.
The experimental uncertainties originate from two sources, trigger and
reconstruction efficiencies.
A summary of the systematic uncertainties is given in table \ref{tab:syst}.
The individual contributions are discussed below.
\begin{table}[t]
\begin{center}
\begin{tabular}{lr}
\hline\hline
\multicolumn{2}{c}{\bf Experimental uncertainties} \\
\hline
trigger inefficiency & \triggererror{} \\
SpaCal energy scale & \escaleerror{} \\
SpaCal energy resolution & \ereserror{} \\
SpaCal position resolution & \posreserror{} \\
CIP efficiency & \cipefferror{} \\
conversion probability & \cipconverror{} \\
alignment & \alignmenterror{} \\
$z$-vertex distribution & \zvtxerror{} \\
SpaCal cluster finder & \clustererror{} \\
CTD efficiency & \ctdefferror{} \\
LAr energy veto & \larvetoerror{} \\
\hline
& \totalsysexp \\
\hline\hline
\multicolumn{2}{c}{\bf Background uncertainties} \\
\hline
non-elastic QEDC & $1.1\%$ \\ 
elastic DVCS & $0.3\%$ \\
quasi-elastic DVCS & $0.1\%$ \\
diffractive VM production & $0.1\%$ \\
non-resonant diffractive DIS & $0.2\%$ \\
$ep\to ep\,e^{+}e^{-}$ & $0.1\%$ \\
\hline
 & \totalsysbgr\\
\hline\hline
\multicolumn{2}{c}{\bf QEDC theory uncertainties} \\
\hline
higher order corrections & \isrerror{} \\
{\textchanged proton form factor (TPE parametrisation)} & {\textchanged \tpeerror{}} \\
{\textchanged proton form factor (experimental)} & {\textchanged \gegmerror{}} \\
{\textchanged size of generated signal sample} & {\textchanged \mcstaterror{}} \\
\hline
 & \totalsystheor \\
\hline\hline
\end{tabular}
\caption{\label{tab:syst}
  Systematic uncertainties on the determination of the integrated luminosity
  using elastic QEDC events. The different error sources are grouped into three categories: experimental, background related and theory related uncertainties.
The error sources are described in detail in the
  text.}
\end{center}
\end{table}
Additional time-dependent uncertainties are present in cases where the
integrated luminosity determined in the present analysis is applied to
subsets of the H1 data.

\subsection{Trigger uncertainties}

The main trigger condition is based on calorimetric information in
the SpaCal.
It has an efficiency of more than $95\%$ for clusters with
energies $E>6\,\text{GeV}$, rising to above $99.8\%$ for energies
$E>10\,\text{GeV}$. These efficiencies are verified using independent
trigger conditions for a selection of DIS events with the scattered
electron in the SpaCal. 
Both the electron and the photon from the elastic QEDC reaction
create clear signals above the trigger condition thresholds, hence the trigger inefficiency on the
QEDC selection is negligible.
However, for certain time periods there were small regions opposite in
$\varphi$ with reduced trigger efficiency.
This leads to an uncertainty of $0.02\%$.
The other trigger conditions are related to timing signals from the VETO
system and from the SpaCal calorimeter, designed to veto non $ep$
background. 
These trigger conditions in conjunction with the varying HERA beam
conditions cause inefficiencies of typically $1\%$ for data taken up
to the year 2005 and of typically $0.2\%$ afterwards. These
inefficiencies mainly originate from particles from beam related background recorded within
genuine $ep$ collision events.
%
The veto inefficiencies are corrected for by applying time-dependent
weights to the data events.
The corresponding systematic uncertainty is \triggererror{}.

\subsection{Reconstruction uncertainties}

Reconstruction uncertainties originate mainly from the understanding of the
SpaCal response to electrons and photons.
The primary SpaCal energy calibration is done using electrons in DIS
events \cite{Collaboration:2010ry}.
It corrects for time-dependent or spatial non-uniformities of the
calorimeter response.
However, the response of the SpaCal is slightly different for electrons and
photons, mostly due to the presence of dead material in front of the
calorimeter and due to final state radiation of the electrons.
Furthermore, it is found that the primary calibration can be improved by
correcting the energy response as a function of the transverse cluster
size, $R_{\log{}}$.
For the QEDC analysis, multiplicative calibration factors are
applied to the SpaCal cluster energies for electrons, non-converted photons
and converted photons, respectively.
These factors are taken to have a linear dependence on $R_{\log{}}$.
The corresponding parameters are determined by applying the
double-angle calibration method  
to both the photon and the electron differentially in $R_{\log{}}$.
Distributions of $P_T/P_{T,DA}$, where $P_T$ is the measured
transverse momentum and $P_{T,DA}$ is the predicted transverse momentum,
are investigated for the selection of QEDC events with the cut on the
momentum balance, $\vert \vec{P}_T^{\text{miss}}\vert
<0.3\,\text{GeV}$, relaxed.
The predicted transverse momentum is given by
\cite{Bentvelsen:1992fu,Hoeger:1991wj}
\begin{equation}
P_{T,DA}=2E^0_e\left(\frac{1-\cos\theta_e}{\sin\theta_e}+\frac{1-\cos\theta_\gamma}{\sin\theta_\gamma}\right)^{-1},
\end{equation}
where $\theta_e$ and $\theta_\gamma$ are
the polar angles of the electron and the photon, respectively.
For each $P_T/P_{T,DA}$ distribution, the position of the maximum is
determined in a fit. The calibration parameters are finally determined
from a linear fit as a function of $R_{\log{}}$.
In data (MC), the energy response of the calorimeter to non-converted
photons is found to be on average $3.5\%$ ($2.2\%$) higher than the response to
electrons.
For the event sample of converted photons, the 
energy response to photon candidates is $0.4\%$ lower than the
response to electrons, both for data and MC.
In order to determine the systematic uncertainty, 
the energy scale of electrons and photons is varied separately by $0.5\%$ each.
The size of this variation covers possible systematic effects
originating from the calibration procedure described above.
Furthermore, a simultaneous variation of the electron and photon
energy scale by another $0.5\%$ is considered as systematic
uncertainty, originating from the primary energy
calibration\cite{Collaboration:2010ry}.
In total, all energy scale variations together contribute to the
uncertainty on ${\cal L}^{\text{QEDC}}$ by \escaleerror{}.

\begin{figure}[htbp]
  \begin{center}
\includegraphics[width=0.5\figwidth]{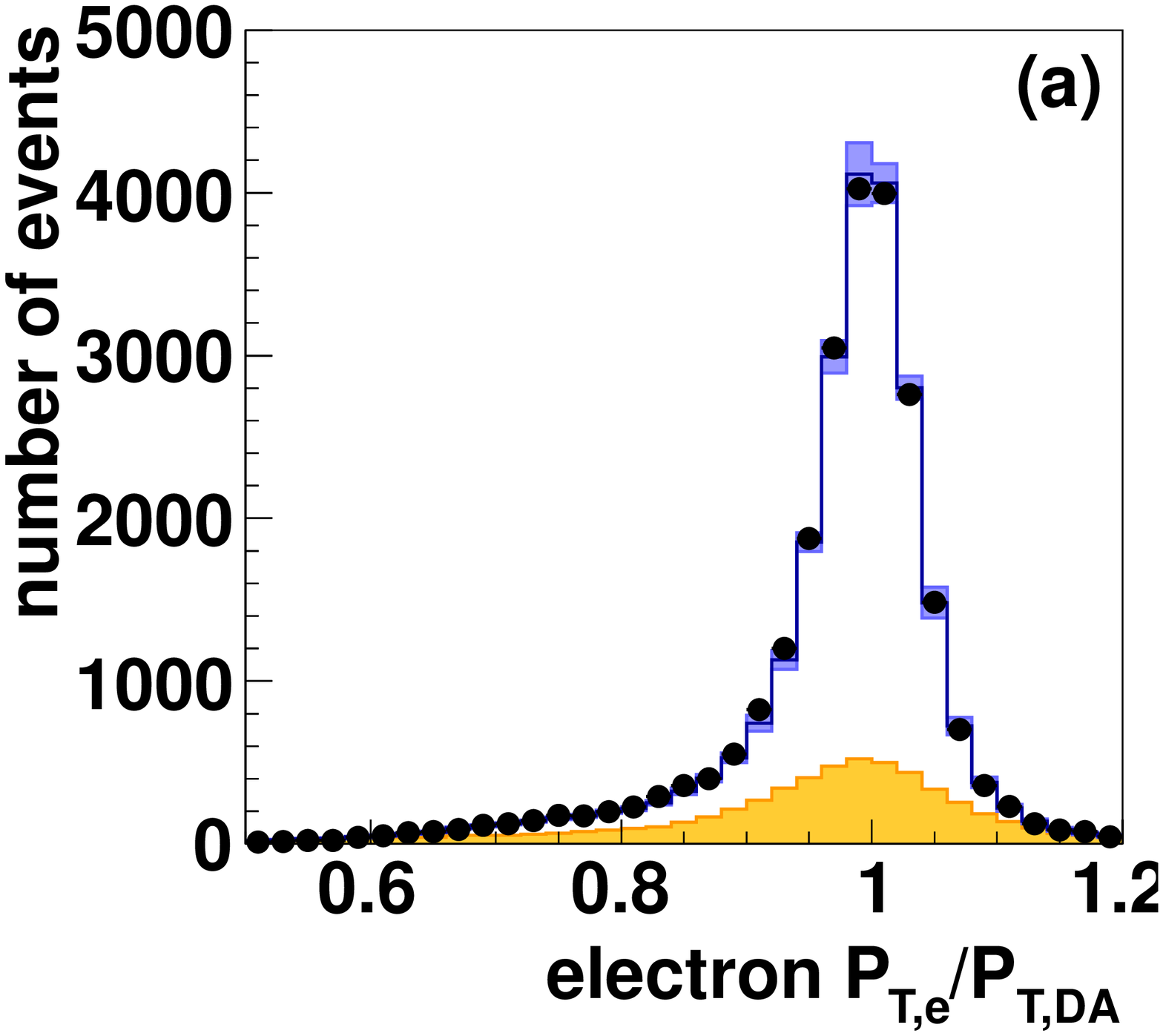}
\includegraphics[width=0.5\figwidth]{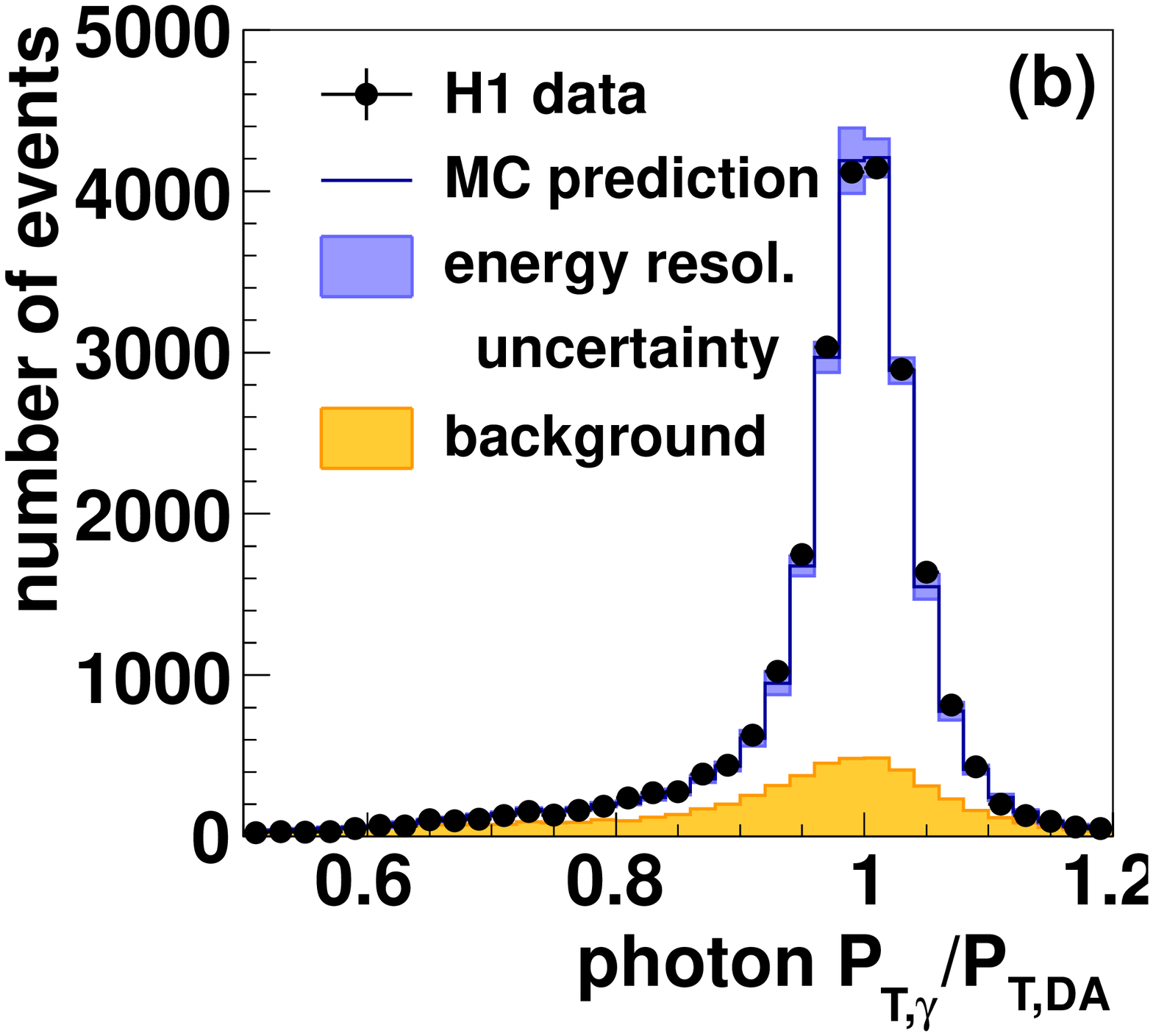}
\caption{\label{fig:calibration}
  Distributions of the ratio of measured to predicted transverse
  momentum for (a) electrons and (b) photons. The predicted transverse
  momentum $P_{T,DA}$ is calculated 
  using the double angle method. The data are
  shown as black dots. The simulation including background, normalised
  to the integrated 
  luminosity as determined in this analysis, is shown as a solid
  line, with the systematic uncertainty originating from the limited
  knowledge of the energy resolution attached as a shaded area.
  Also shown is the contribution from background processes.}
\end{center}
\end{figure}

In addition to the calibration factors, the energy resolution is 
determined from fits to the $P_T/P_{T,DA}$ distributions, however without
dividing the sample into bins of $R_{\log{}}$.
Figure \ref{fig:calibration} shows the distributions of $P_{T,e}/P_{T,DA}$ and
$P_{T,\gamma}/P_{T,DA}$ for electrons and photons, respectively.
The distributions are peaked at $1$, as expected after 
calibration.
Near the peak, the distribution is more asymmetric
towards smaller energies for electrons as compared to photons.
This is attributed to final state radiation and energy losses
in the material located in front of the calorimeter.
The original MC simulation (not shown in figure \ref{fig:calibration})
has deficits to describe both widths and tails towards lower
transverse momenta. The effect exists for both electrons and
photons.
It is corrected for by applying an extra smearing of the reconstructed
energies in the MC simulation.
An energy offset $\Delta E=(\delta-\tau)\,E^0_e/2$ is subtracted, where $\delta$
is a random number drawn from an exponential distribution, i.e.~the 
probability density to find $\delta>0$ is given by
$f(\delta)=1/\tau\exp[-\delta/\tau]$.
By construction, $\Delta E\sim (\delta-\tau)$ has an expectation value of
$\langle\Delta E\rangle=0$.
This has the desired effect that the peak position of $P_T/P_{T,DA}$ is
affected only little by the smearing.
Two independent parameters $\tau_e$ and $\tau_\gamma$ are
foreseen to describe the expectation values of the exponential
probability distributions for electrons and photons, respectively.
It turns out that both $\tau_e$ and $\tau_\gamma$ take the same
central value, $\tau_e=\tau_\gamma=0.010$.
Figure \ref{fig:calibration}a shows that the distribution of
$P_{T,e}/P_{T,DA}$ is described by the smeared simulation within a variation
of the smearing parameter $\tau_e=0.010\pm0.005$.
Similarly, for photons, $P_{T,\gamma}/P_{T,DA}$ (Figure
\ref{fig:calibration}b) is described within the variation
$\tau_\gamma=0.010\pm0.005$.
The  $\tau_e$ and $\tau_\gamma$ variations together cause an
uncertainty on ${\cal L}^{\text{QEDC}}$ of \ereserror{}.

\begin{figure}[htbp]
  \begin{center}
\includegraphics[width=0.5\figwidth]{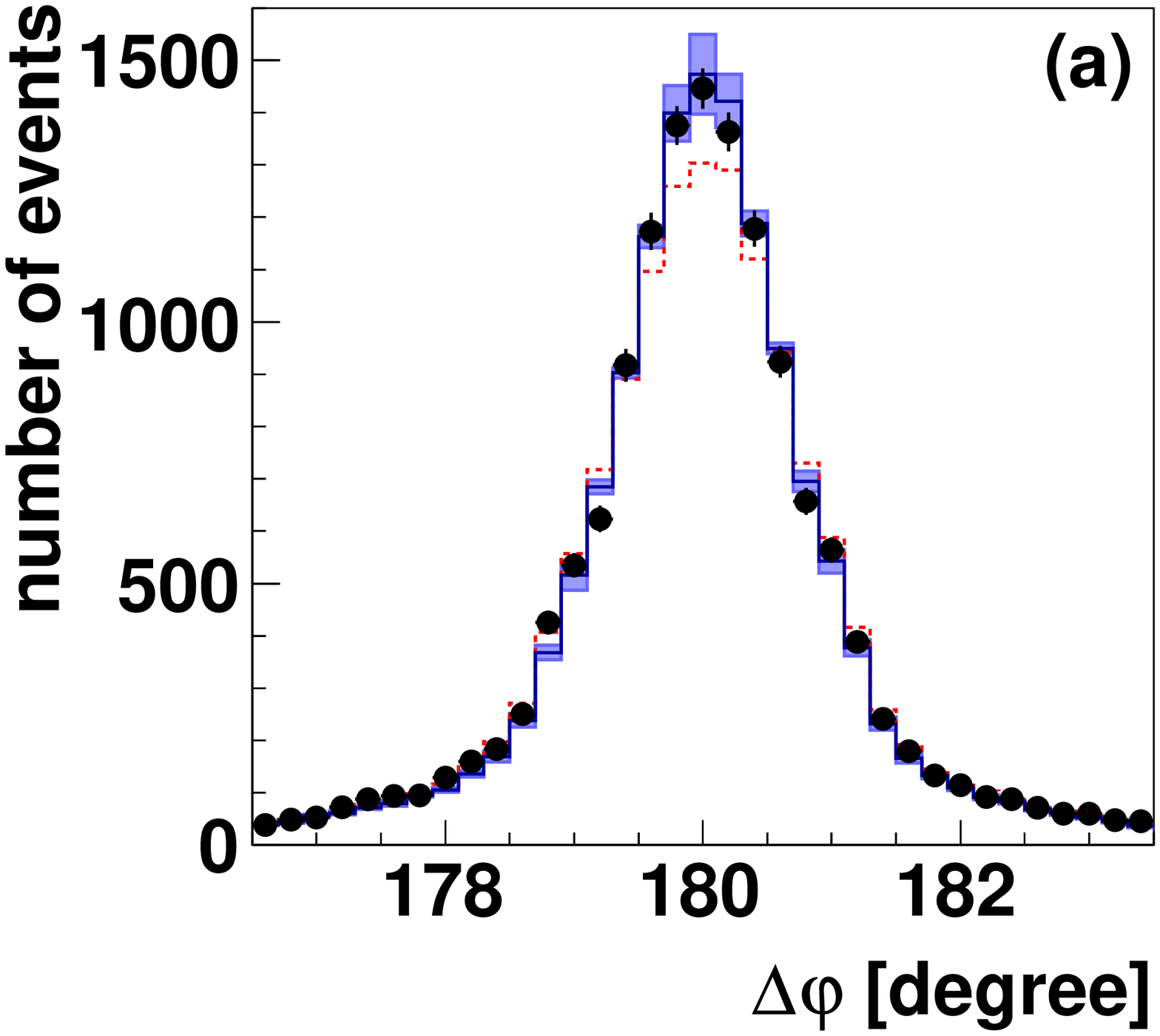}
\includegraphics[width=0.5\figwidth]{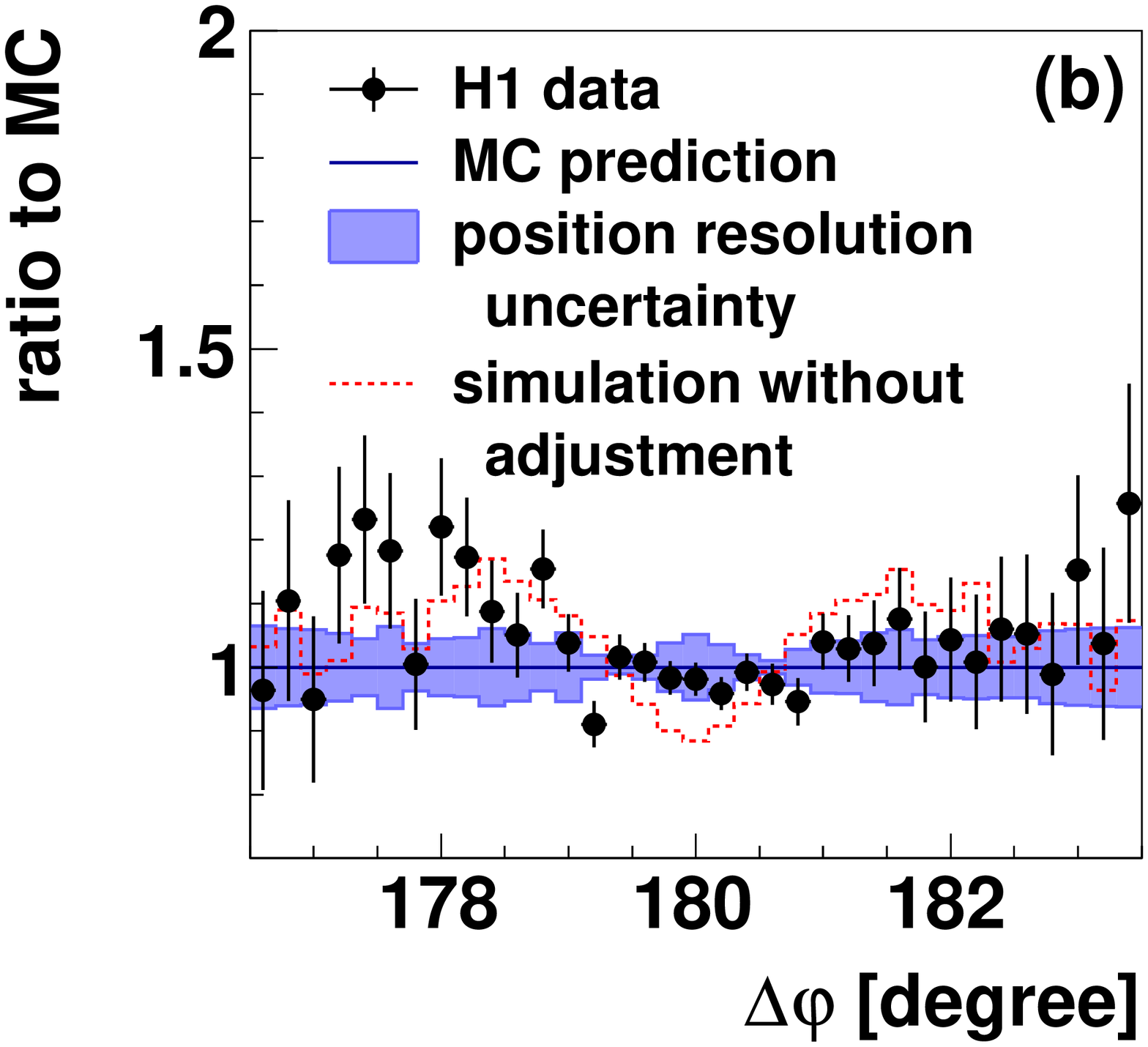}
\caption{\label{fig:posres}
  Distributions of the difference in azimuth between the electron and
  the photon. In 
  (a) the event counts are shown, whereas in (b) the ratio to the 
  expectation is drawn. The data are shown as black dots with
  statistical uncertainties indicated as vertical bars.
  The simulation including background, normalised to the integrated
  luminosity as determined
  in this analysis, is shown as a solid line with the systematic
  uncertainty originating from the limited knowledge of the position
  resolution attached as a shaded area. The distribution 
  predicted by the simulation prior to adjusting the position
  resolution is shown by the dashed line.}
\end{center}
\end{figure}

The SpaCal cluster position resolution in MC is worse than in data.
This effect has been identified using the difference in azimuth of the
electron and the photon, $\Delta\varphi$, shown in figure \ref{fig:posres}.
The original simulation has a deficit at $\Delta\varphi$ near $180^{\circ}$,
corresponding to a resolution worse than in data.
In order to improve the description of data by MC, the reconstructed
cluster positions in MC, $\vec{x}_{\text{rec}}$, are modified such that they are
closer to the extrapolated SpaCal positions of the corresponding generated
particles $\vec{x}_{\text{gen}}$.
For the analysis, the positions
$\vec{x}_{\text{MC}}=(1-f)\vec{x}_{\text{rec}}+f\vec{x}_{\text{gen}}$
are used, where the constant is found to be $f=0.14$. The uncertainty
of $f$ is taken as $0.05$, resulting in an uncertainty on 
${\cal L}^{\text{QEDC}}$ of \posreserror{}.
The data are described by the prediction within that
systematic variation, as demonstrated in figure \ref{fig:posres}.

The CIP efficiency for electrons is determined in data and in the simulation
using DIS events. It is found to be near $99\%$ in data and near
$99.5\%$ for MC, varying as a function of the radial distance of the
SpaCal cluster from the beam, $R$.
A correction as a function of $R$ is made by dropping a
fraction of CIP vertices in the simulation.
The CIP spatial resolution is adjusted using elastic $\rho^0$
production events in DIS. The CIP vertex, reconstructed from the
scattered electron, is compared to the CTD vertex, reconstructed
from the $\pi^{+}\pi^{-}$ pair.
The conversion rate of photons in front of the CIP is underestimated in the
simulation.
In data, the conversion probability is around $32\%$, whereas the MC
predicts $23\%$.
This is corrected by mimicking conversion effects for a fraction of MC
events
with non-converted photons.
For these events, extra CIP clusters near the expected position are
added, and the energy response is scaled to match the expectation for
converted photons.
For estimating systematic effects, the three CIP related corrections described
above are switched off one by one, and the resulting differences on
${\cal L}^{\text{QEDC}}$ are taken as uncertainties. For the CIP
efficiency correction the uncertainty is \cipefferror{}. The CIP
resolution tuning has negligible effect and the conversion probability
leads to an uncertainty of \cipconverror{}.

The alignment of the SpaCal and CIP detectors is done using DIS events.
The interaction vertex is reconstructed using tracks in the CTD,
originating from the hadronic final state.
Using hits in the BPC detector and the energy measured in the SpaCal,
the electron trajectory is extrapolated to the CIP and SpaCal
detectors.
The alignment uncertainties are dominated by the uncertainty on the SpaCal $z$
position. 
Systematic effects are estimated by varying the SpaCal $z$ position by
$\pm 5\,\text{mm}$, resulting in an uncertainty of \alignmenterror{}
on ${\cal L}^{\text{QEDC}}$.

The longitudinal vertex distribution is dominated by a Gaussian near
$z=0$ with a width of approximately $10\,\text{cm}$, as can be seen
in figure \ref{fig:selection}f.
The longitudinal proton beam profile also exhibits prominent satellite peaks
of similar width, leading to collisions in the H1 detector near
$\pm70\,\text{cm}$.
In addition, there is an excess of collisions near $40\,\text{cm}$, as compared
to a simple model which includes only collisions from the main bunch
and from the satellites.
For this analysis, the simulated vertex distribution is re-weighted such that
the full interaction region is described. The difference of
\zvtxerror{} in ${\cal L}^{\text{QEDC}}$, obtained when using the
simple beam profile model, is taken as a systematic uncertainty.
The reconstructed $z$-vertex distribution after re-weighting is
compared to the data in figure \ref{fig:selection}f. Good agreement is found.
The regions of sizable systematic uncertainty due to the vertex
re-weighting are visible.

The identification of clusters in the SpaCal is
checked by relaxing the $R_{\log{}}<6\,\text{cm}$ condition. Removing the
$R_{\log{}}$ condition results in a somewhat smaller number of selected QEDC
events, because a third SpaCal cluster is accepted more often, 
leading to the rejection of the corresponding events.
%
This procedure leads to a change in 
${\cal L}^{\text{QEDC}}$ of \clustererror{} which is
considered as systematic uncertainty related to the cut in $R_{\log{}}$.
The uncertainty on the CTD track reconstruction efficiency of
typically $2\%$ per track affects the analysis through the track
veto, resulting in an uncertainty of \ctdefferror{} on ${\cal L}^{\text{QEDC}}$.
The veto on the energy in the forward part of the LAr calorimeter
is checked by relaxing the veto condition to $E<1\,\text{GeV}$, resulting in
an uncertainty of \larvetoerror{} on ${\cal L}^{\text{QEDC}}$.

\subsection{Background uncertainties}

\begin{figure}[htbp]
  \begin{center}
\includegraphics[width=0.5\figwidth]{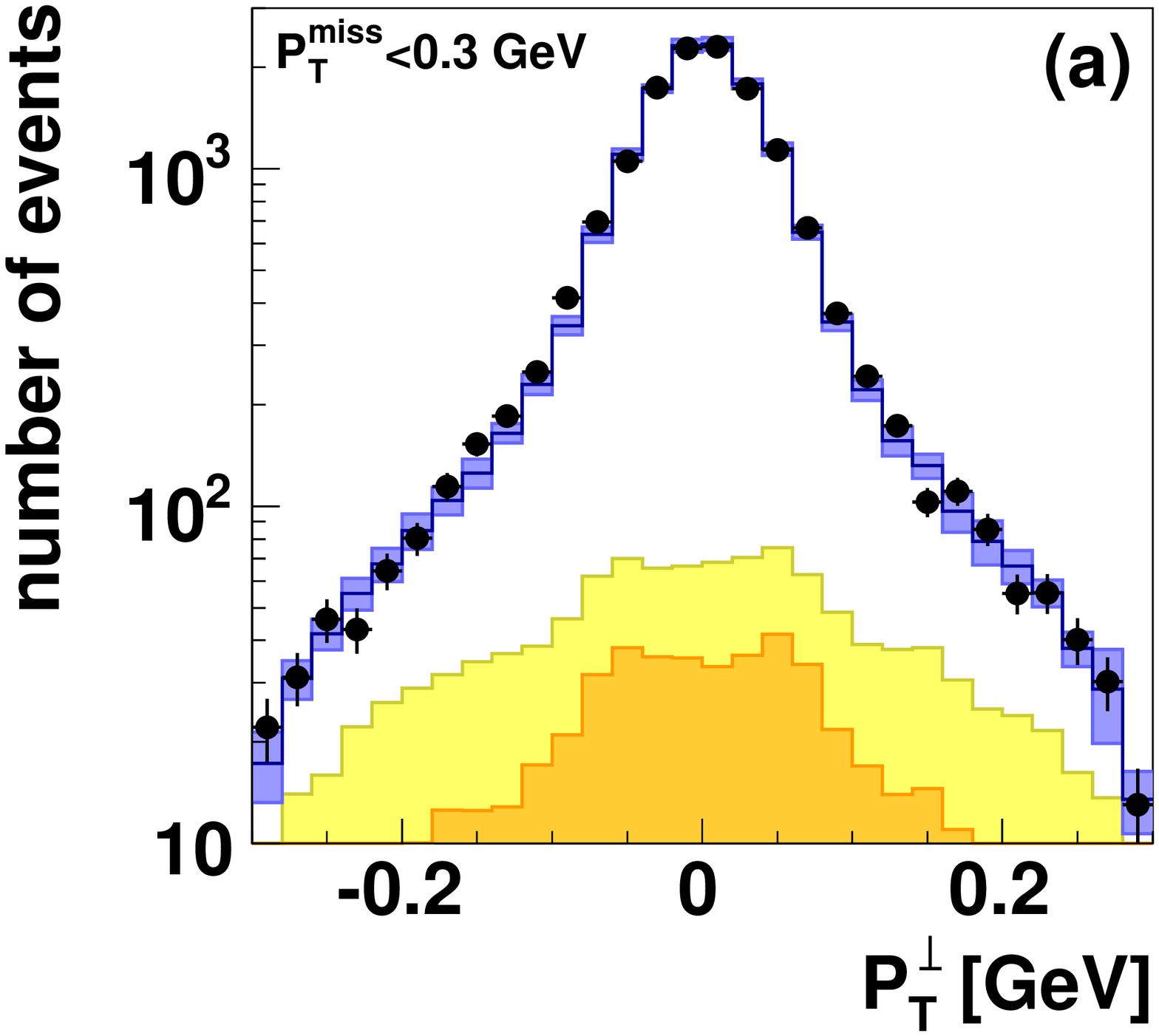}
\includegraphics[width=0.5\figwidth]{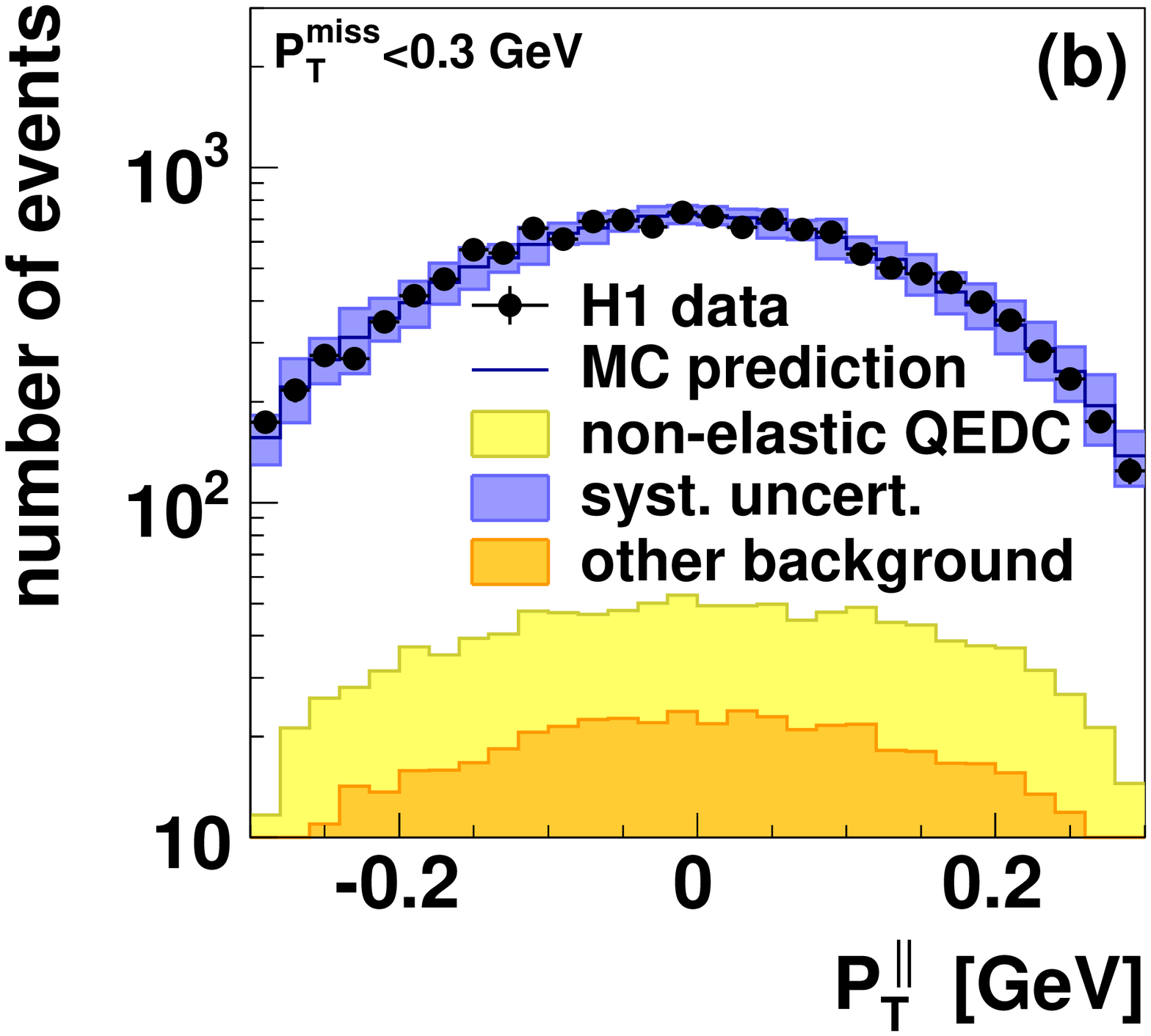}
\includegraphics[width=0.5\figwidth]{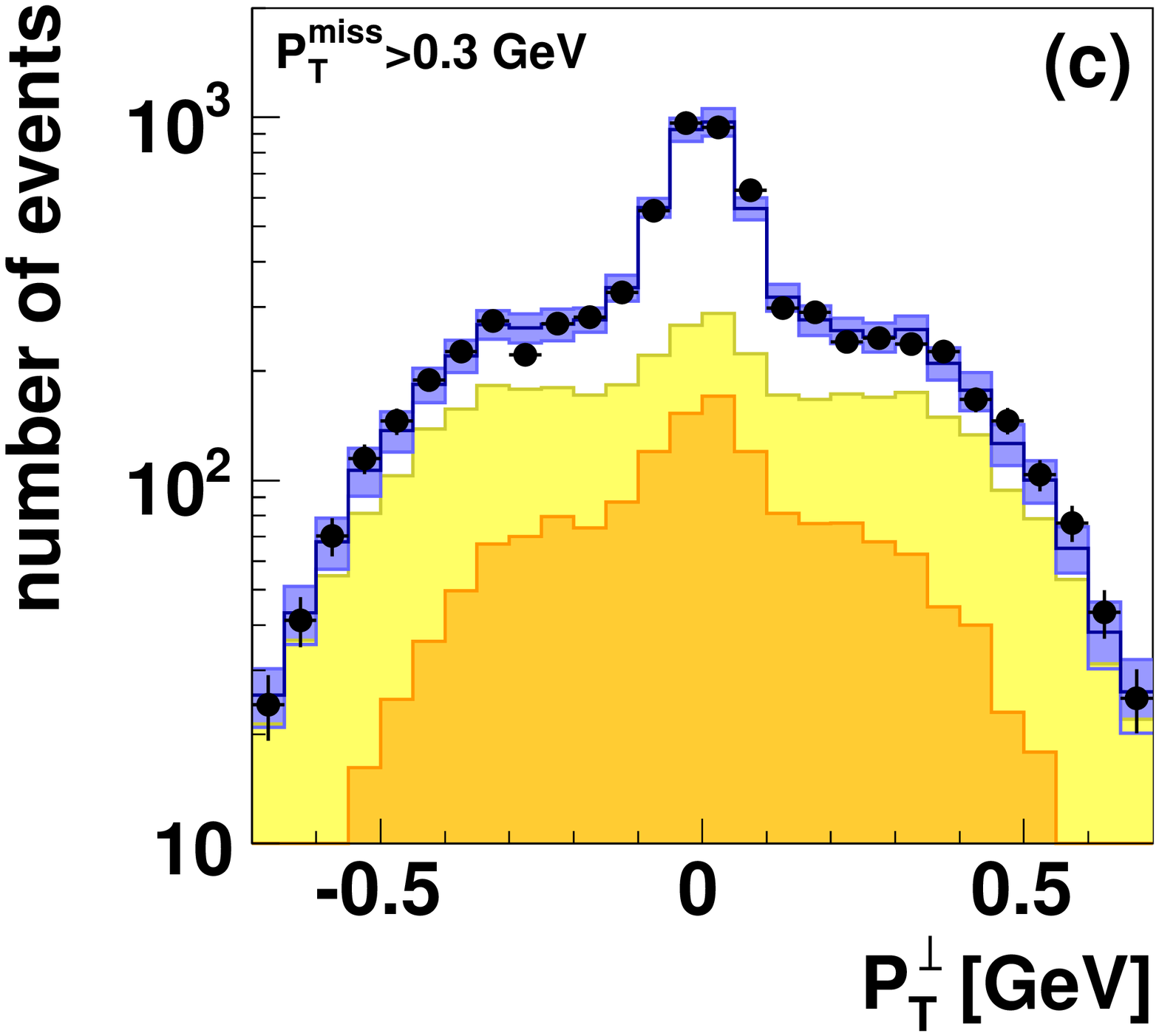}
\includegraphics[width=0.5\figwidth]{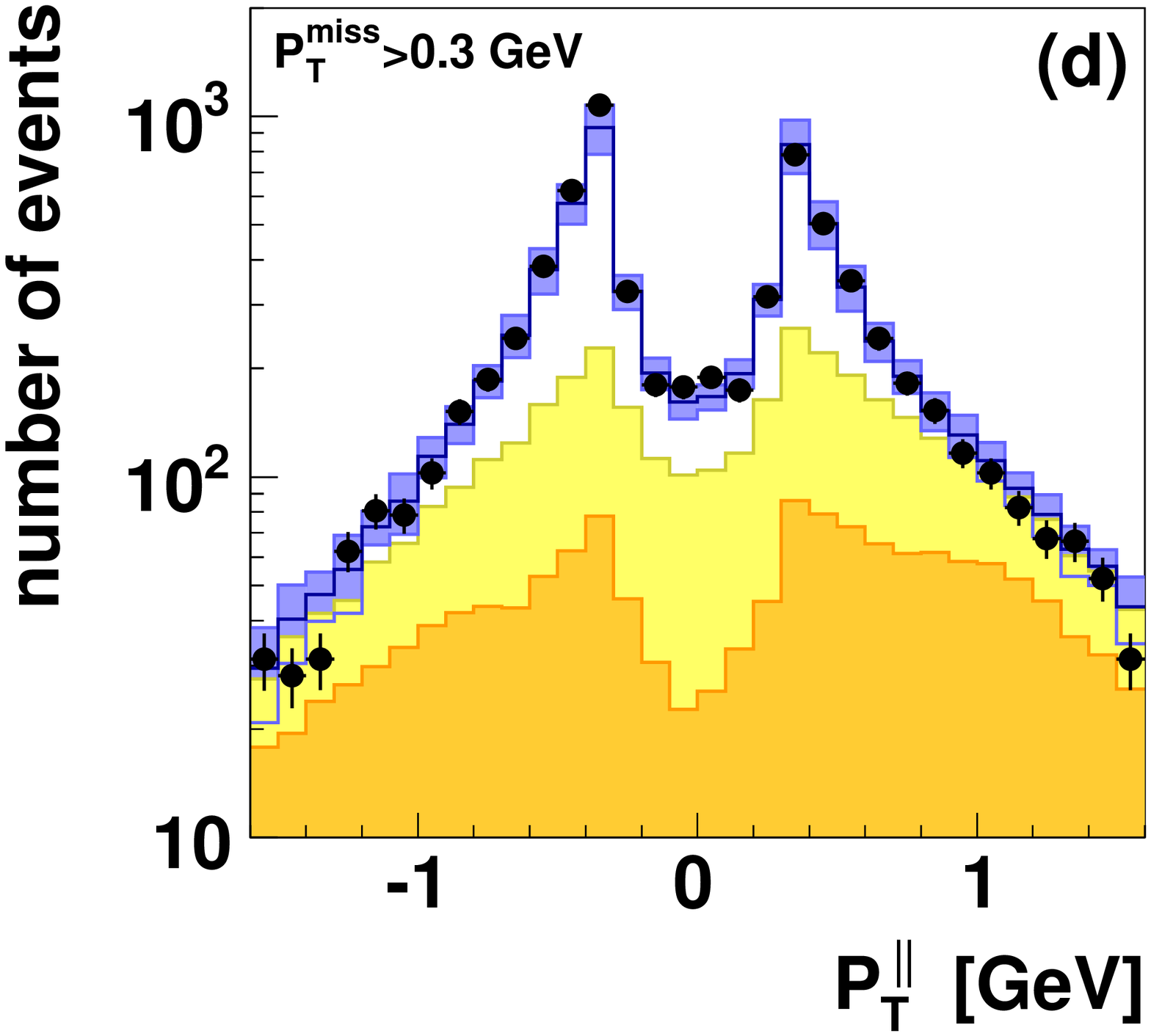}
\caption{\label{fig:ptcomponents}
  Distributions of components of the photon plus electron transverse
  momentum sum, 
  $\vec{P}_T^{\text{sum}}$: (a) and (c) the component perpendicular to the electron
  transverse momentum, (b) and (d) the component parallel to the electron
  transverse momentum. The upper row, (a) and (b) shows the distributions
  inside the analysis phase space, the lower row (c) and (d) shows the
  distributions for $\vert\vec{P}_T^{\text{miss}}\vert>0.3\,\text{GeV}$.
  The data are shown as black
  dots with the statistical uncertainties indicated as vertical
  bars. The simulation including background, 
  normalised to the integrated luminosity
  as determined in this analysis, is
  indicated as a solid line, with the  systematic uncertainties
  attached as shaded area. Also shown are
  the contributions from non-elastic QEDC and from other background sources.}
\end{center}
\end{figure}

The normalisation of quasi-elastic and inelastic QEDC events predicted by the
COMPTON22 event generator depends mainly on parameters related to the nucleon
excitation and on the proton structure function parametrisation at low momentum
transfer, respectively.
These parameters are not known very well.
For this reason, the normalisation of the sum of 
these contributions, referred to as ``non-elastic QEDC'', is verified by
investigating the vector sum of the electron and photon transverse momenta,
$\vec{P}_T^{\text{sum}}=-\vec{P}_T^{\text{miss}}$.
The vector
$\vec{P}_T^{\text{sum}}$ is decomposed into components parallel to
($P_T^{\parallel}$) and perpendicular to ($P_T^{\perp}$) the electron
transverse momentum.
The distributions of $P_T^{\parallel}$ and $P_T^{\perp}$ are shown in
figure \ref{fig:ptcomponents} inside the analysis
phase space as well as for
$\vert\vec{P}_T^{\text{miss}}\vert>0.3\,\text{GeV}$ with all other
selection criteria applied.
Both the parallel and the perpendicular components are described well.
Outside the nominal analysis phase space the non-elastic QEDC
contributions dominate at large  $P_T^{\parallel}$ or at large
$P_T^{\perp}$.
The $P_T^{\parallel}$ distribution is somewhat asymmetric for
$\vert\vec{P}_T^{\text{miss}}\vert>0.3\,\text{GeV}$, because in
contrast to photons the SpaCal response to electrons has tails
towards low energies, as discussed in the previous section.
The normalisation of the non-elastic QEDC contribution is tested by performing
fits to either $P_T^{\parallel}$ or $P_T^{\perp}$ for
$\vert\vec{P}_T^{\text{miss}}\vert>0.3\,\text{GeV}$.
The normalisation factors observed in these fits are
compatible with the COMPTON22 prediction within $25\%$, which is used
as normalisation uncertainty for the non-elastic QEDC processes.

The DVCS cross section predictions obtained with the MILOU program are in
agreement with recent H1 measurements \cite{:2009vda}.
Uncertainties of $20\%$ for the elastic DVCS process and $50\%$
for proton dissociative DVCS are considered.

The elastic VM production rates are normalised using dedicated selections as
close as possible to the QEDC analysis.
However, instead of requiring a photon in the SpaCal, a vector meson is
reconstructed.
The decays $\rho^0\to\pi^{+}\pi^{-}$, $\phi\to K^{+}K^{-}$, $J/\psi\to
\ell^{+}\ell^{-}$ ($\ell=e,\mu$) and $\psi^{\prime}\to
\ell^{+}\ell^{-}$ are reconstructed from
two oppositely charged tracks, detected in the CTD.
The decay $\omega\to\pi^{+}\pi^{-}\pi^{0}$ is reconstructed from two
charged tracks and one or two neutral calorimeter clusters.
The decay $\Upsilon\to e^{+}e^{-}$ is reconstructed using a sample of
photoproduction events, where an $e^{+}e^{-}$ pair from the
$\Upsilon$ decay is reconstructed in the SpaCal, one of the SpaCal
clusters matched with a central track, and the scattered electron is
outside the acceptance of the H1 detector. The following normalisation
uncertainties are found: $20\%$ on $\rho^0$ and $\phi$ production, $50\%$
on $J/\psi$, and $100\%$ on $\omega$, $\psi^{\prime}$ and
$\Upsilon$. Possible contributions from $\rho(1450)$ production are
covered by the $\rho^0$ normalisation uncertainty.

The rate of non-resonant diffractive events, simulated using RAPGAP, is
normalised using a selection of low multiplicity final states, where the
electron is reconstructed in the SpaCal and one up to three additional
charged or neutral particles are found. The uncertainty is estimated
to be $30\%$.

For the QED processes modelled by GRAPE, an uncertainty of $10\%$ is assumed,
taking into account possible higher order effects.

{\textchanged The uncertainties on the various background samples originating from
the finite sizes of the generated event samples are negligible as
compared to the uncertainties discussed above.}

\subsection{QEDC theory uncertainties}
\label{text:theoryerror}

Uncertainties to the elastic QEDC cross section arise mainly from two sources:
higher order corrections and the knowledge of the proton form factors.

In the original COMPTON22 event generator, higher orders are simulated in the
peaking approximation \cite{peakingisr}.
Improved higher order corrections have been calculated
\cite{Anlauf:1991vi} 
using a photon radiator \cite{Montagna:1991ku}.
For the purpose of this analysis, the COMPTON22 events are assigned weights,
determined such that the cross section predicted by the photon radiator
method is reproduced. 
The difference {\textchanged of \isrerror{}} to the original COMPTON22 prediction is taken as
systematic uncertainty due to higher order effects.
The elastic QEDC cross section also depends on the proton electric and
magnetic form factors.
{\textchanged In the original COMPTON22 generator, only a simple parametrisation of
the form factors, using one parameter, is implemented.
For this analysis, recent form-factor fits of elastic $ep$ scattering data
\cite{Arrington:2007ux} are taken into account, using an event weighting
technique.
The form-factor parameterisations \cite{Arrington:2007ux} are corrected for the
effects of two-photon exchange (TPE), but parameterisations not
including TPE corrections are also provided.
In the COMPTON22 generator, TPE contributions are not included when
calculating the cross-section.
For this reason the form factor parameterisations not including TPE
corrections are used to calculate elastic QEDC cross sections
\cite{arringtonprivate}.
When computing the COMPTON22 cross section with the TPE corrected form
factors, the analysis result changes by \tpeerror{}. This difference
is included as systematic uncertainty.
Experimental uncertainties on the form factor parametrisation are also
considered.
Such uncertainties are available with the parametrisation
\cite{Venkat:2010by} which includes TPE corrections.
It has been verified that  the difference between \cite{Venkat:2010by}
and \cite{Arrington:2007ux} is completely negligible for the
purpose of this analysis if the TPE corrections are included.
The experimental uncertainties on the form factors $G_E(\vert t\vert)$ and
$G_M(\vert t\vert)$ approach zero for
$\vert t\vert\to 0$, because the parameterisations enforce $G_E(0)=1$
and $G_M(0)=\mu_p$.
At $\vert t\vert=0.007$, which is the average momentum transfer predicted for
the events selected in this analysis, the uncertainties on $G_E$
($G_M$) are $0.1\%$ ($0.2\%$).
The elastic QEDC cross section at fixed $\vert t\vert$ is given by a 
linear combination of $G_E^2(\vert t\vert)$ and $G_M^2(\vert t\vert)$,
where the $G_E^2$ contribution dominates at small $\vert t \vert$.
When propagating the $\vert t\vert$ dependent parametrisation
uncertainties on $G_E$ and $G_M$ to the luminosity measurement, 
an uncertainty of \gegmerror{} is found.
In addition to the theory uncertainties related to higher order
corrections and proton form factors, discussed above, there is a
statistical uncertainty on the theory prediction originating from the
finite size of the generated signal event sample.
It amounts to \mcstaterror{}.}

\begin{figure}[htbp]
  \begin{center}
\includegraphics[width=0.5\figwidth]{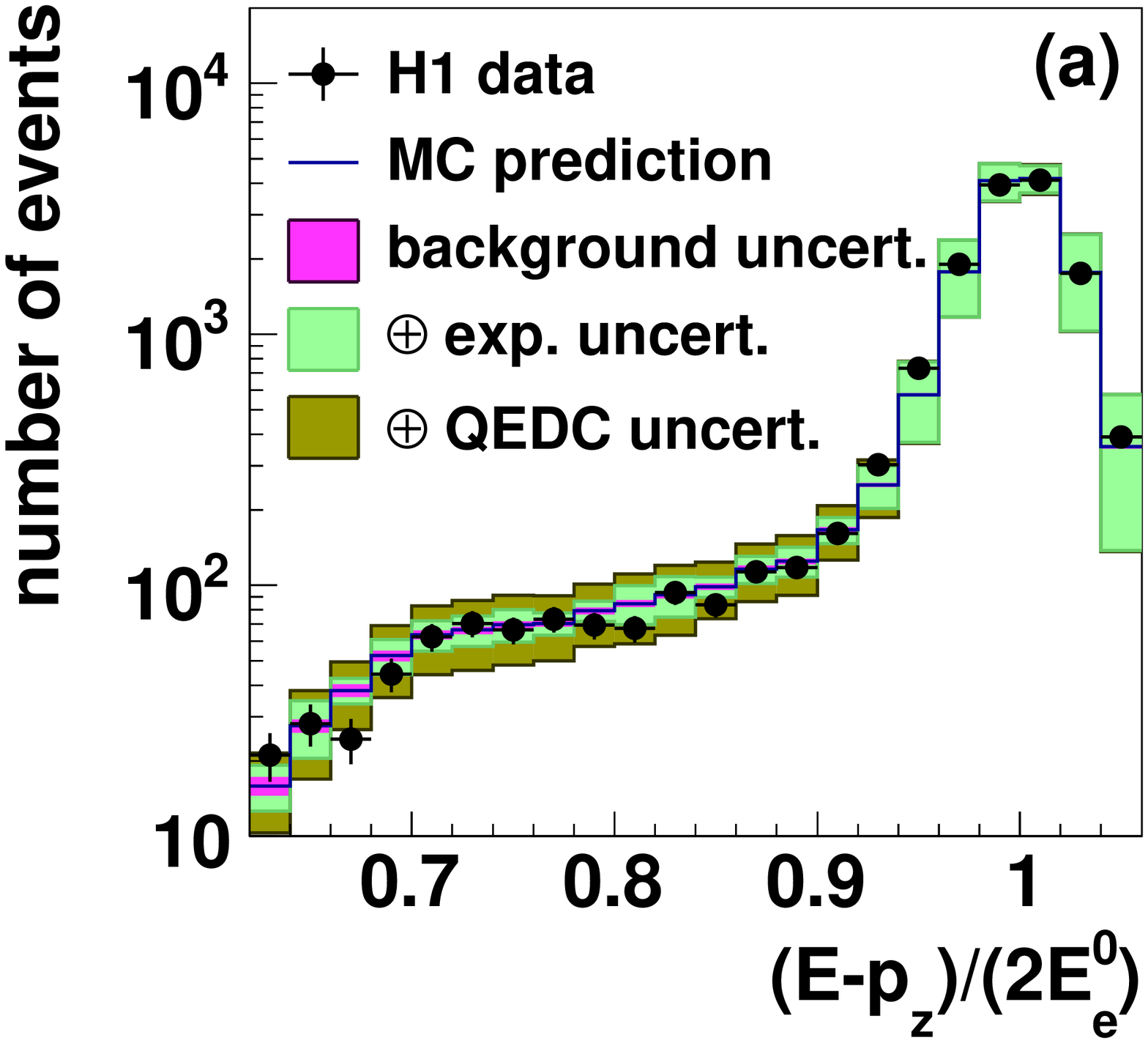}
\includegraphics[width=0.5\figwidth]{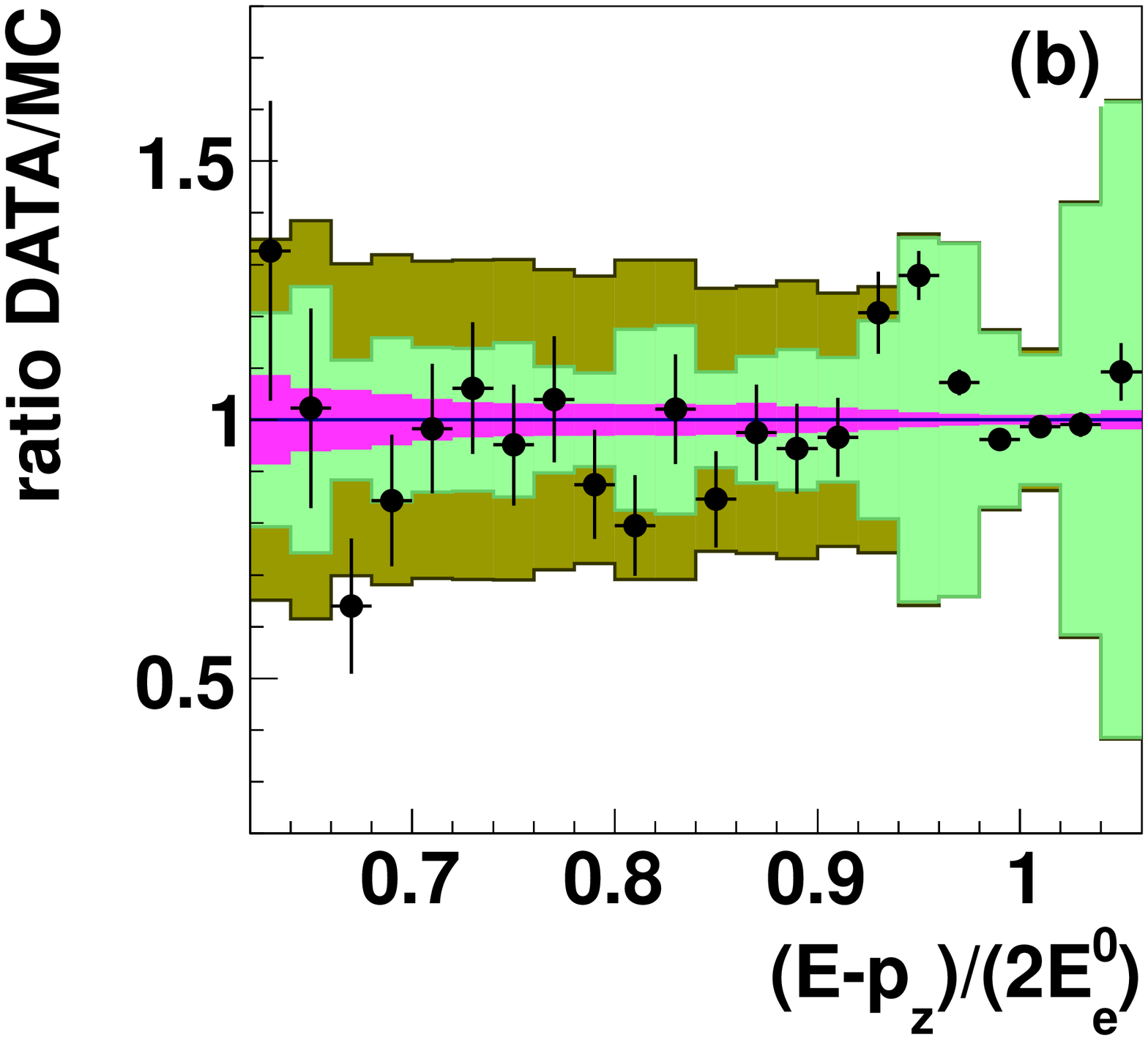}
\caption{\label{fig:epz}
  Distribution of the variable $(E-p_z)/{2E^{e}_0}$ calculated
  from the sum of the electron and 
  photon four-momenta. In (a) the event counts are shown, whereas in
  (b) the ratio of data to expectation is drawn.
  The data are shown as black dots with the statistical uncertainties
  indicated as vertical bars. The simulation including background,
  normalised to
  the integrated luminosity as determined in this analysis,
  is indicated as a solid
  line, with various components of the systematic uncertainties attached as
  shaded areas.}
\end{center}
\end{figure}

In figure \ref{fig:epz} the distribution of the variable $(E-p_z)/(2E^0_e)$ is
studied.
This variable is calculated from the sum of the four-momenta of the
electron and the photon.
The distribution of this variable is expected to peak at $1$.
The tail to small $(E-p_z)/(2E^0_e)$ originates from initial state
radiation, whereas values larger than $1$ show up due to resolution effects. 
The data are described within the systematic and
statistical uncertainties for both small and large values of $(E-p_z)/(2E^0_e)$. As expected, the
peak region is dominated by experimental uncertainties, whereas the region of 
small $(E-p_z)/(2E^0_e)$ is dominated by uncertainties of the QEDC cross
section.

\subsection{Time-dependent uncertainties}

\begin{figure}[htbp]
  \begin{center}
\includegraphics[width=\figwidth]{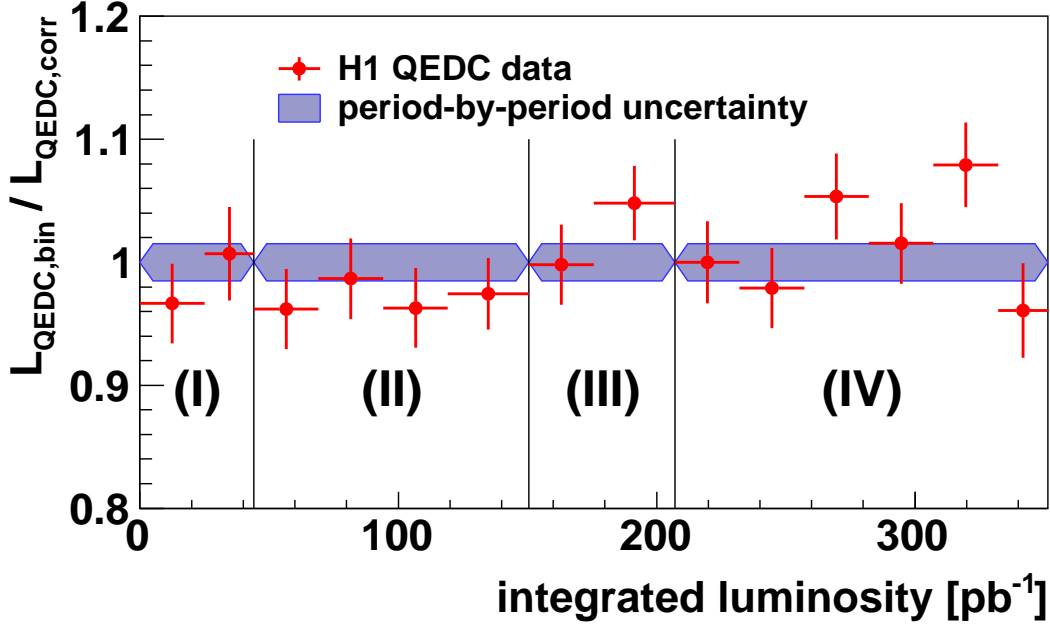}
\caption{\label{fig:yield}
  Integrated luminosity measured from elastic QEDC
  events in bins of approximately $25\,\text{pb}^{-1}$, divided by the
  integrated luminosity derived from the QEDC analysis on the full sample with
  time-dependent corrections applied. The statistical uncertainties of
  the binned QEDC analysis as well as the uncertainties of the
  time-dependent corrections, here applied to four data taking periods
  (I)--(IV), are indicated.}
\end{center}
\end{figure}

In order to apply the integrated luminosity ${\cal L}^{\text{QEDC}}$ to other
physics analyses, possibly using time restricted H1 data sets, a luminosity
calculation differential in time is required.
This is achieved using DIS events measured in the SpaCal. The DIS
selection follows the selection described in
\cite{Collaboration:2010ry} but is restricted in electron polar angle
to the range $167^{\circ}<\theta_e<172^{\circ}$ such that the expected
event yield is most insensitive to changing beam conditions, in particular
to the average position of the interaction vertex in $z$.
In addition, the electron energy range is restricted to
$15<E<25\,\text{GeV}$ and the electron transverse cluster size to
$R_{\log{}}<4.5\,\text{cm}$. The longitudinal vertex position,
measured in the CTD, is restricted to be within $\pm 35\,\text{cm}$
around the nominal interaction point.
The DIS event counts for each run\footnote{H1 data is grouped into
  runs, where new runs are started whenever data taking conditions
  changed. A run typically spans about $30$ minutes of data,
  equivalent to an integrated luminosity of about 
  $30\,\text{nb}^{-1}$.} are used to define relative integrated luminosities
of the runs, and the overall normalisation is taken from the QEDC
analysis.
The statistical uncertainty of the DIS selection is negligible, but it has a
time-dependent systematic error of $1.5\%$.
This uncertainty originates mainly from the SpaCal trigger and vertex
reconstruction efficiencies \cite{Collaboration:2010ry}.
Figure \ref{fig:yield} shows the results of the elastic QEDC
analysis performed in bins of about $25\,\text{pb}^{-1}$, normalised
to the global QEDC analysis with the DIS yield corrections applied.
Four data taking periods, corresponding to distinct configurations of
the HERA machine or the H1 detector, are indicated.
The HERA machine has been operated with $e^{+}p$ beams for periods (I) and
(IV) and with $e^{-}p$ beams for period (II) and (III).
The H1 detector has been opened for the repair or upgrade of various
components between period (I) and (II) as well as between period (II) and (III).
The two methods of measuring differential in time  are in good
agreement, taking into account the statistical fluctuations of the
time-dependent QEDC analysis and the time-dependent systematic
uncertainties of the DIS yield method.

\section{Results}

The integrated luminosity of the data collected in the years
2003 to 2007 is determined using elastic QED
Compton events.
For the data sample as used in this paper, an integrated luminosity of
${\cal L}^{\text{QEDC}}=351.6\pm 8.0\,\text{pb}^{-1}$ is measured.
The statistical uncertainty amounts to \totalerrstat{}, whereas the
total systematic error is \totalerrsys{}.
The integrated luminosity is in agreement with
the Bethe-Heitler measurement, ${\cal L}^{\text{BH}}=\bhluminosity$.
The corrections needed to measure the integrated luminosity of
arbitrary time restricted data samples, such as samples comprising
only $e^{+}p$ or only $e^{-}p$ beams, induce a further uncertainty of $1.5\%$.

\section{Conclusions}

The elastic QED Compton process is used to determine the integrated
luminosity of the H1 data taken in the years 2003 to 2007.
The systematic uncertainties are about equally shared
between experimental uncertainties, understanding of the elastic QEDC
cross section and understanding of the background to the measurement. The
statistical uncertainty is small compared to the systematic uncertainties.
The new measurement method presented in this paper allow a determination of the integrated luminosity with a precision of \totalerr{}.

\section*{Acknowledgements}

We are grateful to the HERA machine group whose outstanding
efforts have made this experiment possible. 
We thank the engineers and technicians for their work in constructing and
maintaining the H1 detector, our funding agencies for 
financial support, the
DESY technical staff for continual assistance
and the DESY directorate for support and for the
hospitality which they extend to the non DESY 
members of the collaboration.
{\textchanged We would like to thank J.~Arrington for useful discussions and for providing 
calculations of uncertainties on the proton form factor parameterisations.}


\pagebreak

\end{document}